\pdfoutput=1
\documentclass[11pt,a4paper]{article}
\usepackage{jheppub}
\usepackage{subfig}
\usepackage{graphicx}
\usepackage{hyperref}
\usepackage{multirow}
\usepackage{color}

\begin{document}

\newcommand{\beq}{\begin{eqnarray}}
\newcommand{\eeq}{\end{eqnarray}}
\newcommand{\non}{\nonumber\\}
\newcommand{\D}{\mathcal{D}}
\newcommand{\p}{\partial}
\newcommand{\Tr}{{\rm Tr}}
\newcommand{\diag}{{\rm diag}}
\newcommand{\sign}{{\rm sign}}
\newcommand{\sech}{{\rm sech}}
\newcommand{\arctanh}{{\rm arctanh}}
\newcommand{\arccot}{\mathop{\rm arccot}}

\title{Gravitating BPS Skyrmions}

\author{Sven Bjarke Gudnason,${}^{1}$}
\author{Muneto Nitta${}^2$ and}
\author{Nobuyuki Sawado${}^3$}
\affiliation{${}^1$Institute of Modern Physics, Chinese Academy of Sciences,
  Lanzhou 730000, China}
\affiliation{${}^2$Department of Physics, and Research and
    Education Center for Natural Sciences, Keio University, Hiyoshi
    4-1-1, Yokohama, Kanagawa 223-8521, Japan}
\affiliation{${}^3$Department of Physics, Tokyo University of Science, Noda,
  Chiba 278-8510, Japan}
\emailAdd{bjarke(at)impcas.ac.cn}
\emailAdd{nitta(at)phys-h.keio.ac.jp}
\emailAdd{sawado(at)ph.noda.tus.ac.jp}

\abstract{
The BPS Skyrme model has many exact analytic solutions in flat space.
We generalize the model to a curved space or spacetime and find that
the solutions can only be BPS for a constant time-time component of
the metric tensor.
We find exact solutions on the curved spaces: a 3-sphere and a
3-hyperboloid; and we further find an analytic gravitating Skyrmion on
the 3-sphere.
For the case of a nontrivial time-time component of the metric, we
suggest a potential for which we find analytic solutions on
anti-de Sitter and de Sitter spacetimes in the limit of no
gravitational backreaction. We take the gravitational coupling
into account in numerical solutions and show that they are well
approximated by the analytic solutions for weak gravitational
coupling. 
}

\keywords{Skyrmions, solitons, anti-de Sitter, exact solutions}

\maketitle


\section{Introduction}

The Skyrme model, in flat spacetime, was made as an effective field
theory in which baryons are solitons in a mesonic field
theory \cite{Skyrme:1962vh,Skyrme:1961vq}. 
The interest in the Skyrme model increased drastically when it was
shown to describe baryons in large-$N_c$ QCD,
exactly \cite{Witten:1983tw,Witten:1983tx}.
Black holes are conjectured to be characterized only by their mass and
global charges at spatial infinity, which is called the (weak) no-hair
conjecture.
Probably the first (stable) counter example to the no-hair conjecture 
was provided by the black hole with Skyrme hair
\cite{Luckock:1986tr,Droz:1991cx,Heusler:1991xx,Heusler:1992av,Bizon:1992gb}
(see also \cite{Glendenning:1988qy,Volkov:1998cc}).
The Skyrmion black hole solution was also generalized to anti-de
Sitter \cite{Shiiki:2005aq,Shiiki:2005xn} and de
Sitter \cite{Brihaye:2005an} spacetimes. 
Later some works on the late-time evolution of the radiation coming
from a black hole-Skyrmion system was
considered \cite{Zajac:2009am,Zajac:2010mu}.
Gravitating sphalerons in the Einstein-Skyrme system were also
considered recently \cite{Shnir:2015aba}.
After the marriage of general relativity and the Skyrmion, some
studies have put forward potential applications.
In particular, one obvious direction of great interest is the
application of the system to neutron stars
\cite{Piette:2007wd,Nelmes:2011zz,Nelmes:2012uf}. 

Topological solitons on curved spaces (as opposed to curved
spacetimes), have led to important exact solutions in the
literature. A few notable examples are: the Skyrmion on the
3-sphere \cite{Manton:1986pz} and the vortex on the hyperbolic
plane \cite{Witten:1976ck}.

Topological solitons of other types than Skyrmions have been studied
on curved spacetime backgrounds (see e.g.~\cite{Volkov:1998cc}) in the
literature and in particular on spacetimes with nonzero cosmological
constant. It is impossible to make a complete list  
here, but we would like to mention a few works.
The lower-dimensional relative to the Skyrmion, namely the
baby-Skyrmion has been studied on the anti-de Sitter background where
the curvature can mimic a mass term \cite{Elliot-Ripley:2015hoa}.
Monopoles were considered in anti-de Sitter
space, where monopole condensation in the form of a wall in the bulk
can break translational symmetry
spontaneously \cite{Bolognesi:2010nb,Sutcliffe:2011sr}.
Gravitating critically coupled vortices give rise to the
Einstein-Bogomol'nyi equation \cite{Yang:1992}, and in turn to
the conjecture that no coincident vortex solution to latter equation
exists (Yang's conjecture), which was proven only
recently \cite{Alvarez-Consul:2015}.
Gravitating semilocal strings were studied in
Ref.~\cite{Gibbons:1992gt} and non-Abelian strings were coupled to
gravity in Ref.~\cite{Aldrovandi:2007bn}. 

The Skyrme model, although rather phenomenologically successful in
describing nuclei of nature, has one short-coming; namely in its
minimal formulation it gives rise to too large binding energies.
This problem has led to the formulation of a theory with infinitely
many mesons \cite{Sutcliffe:2010et} and to the BPS Skyrme
model\footnote{BPS stands for
Bogomol'nyi-Prasad-Sommerfield and the model is named this way because
the energy is proportional to the topological charge, namely the
baryon charge. }
\cite{Adam:2010fg,Adam:2010ds}.
The BPS Skyrme model consists only of the topological baryon current
(squared) and a potential (so no Skyrme term). The advantages of this
model is that it is integrable and has vanishing binding energies for
all baryon numbers.
Due to its importance in many applications, we choose to study this
model in the context of general relativity in this paper.
We would like to mention a third proposal for decreasing the binding
energies in the Skyrme model, which is based on energy (mass)
bounds \cite{Harland:2013rxa,Adam:2013tga} and a certain repulsive
potential; namely the weakly bound Skyrme
model \cite{Gillard:2015eia}. 

Recently, the BPS Skyrme model has also been used to describe neutron
stars (in analogy with the studies using the normal Skyrme
model) \cite{Adam:2014dqa,Adam:2015lpa}, where the BPS Skyrme model
has a clear advantage. Usually thermodynamic properties are only
obtained after averaging over microscopic quantities in a theory. In
the BPS Skyrme model, however, averaging is not necessary, as the
quantities can be calculated directly from the microscopic fields via
target-space integrals \cite{Adam:2014nba}.
Although many analytic solutions can be found in the BPS Skyrme model
on flat (Minkowski) space, the studies of the BPS Skyrmions coupled to
gravity, required numerical
calculations \cite{Adam:2014dqa,Adam:2015lpa}. 

In this paper we are considering the BPS Skyrme model on curved spaces
and on curved spacetimes and try to find analytic solutions for the
gravitating Skyrmions.
Our first result (in Sec.~\ref{sec:model}) is that in order for the
BPS equation of the BPS 
Skyrme model to solve the second-order equation of motion, the
time-time component of the metric needs to be a constant
($g_{00}={\rm const.}$). Conversely, for a solution to be BPS, it also 
requires the same condition to hold true.
First in Sec.~\ref{sec:curvedspace}, we find analytic solutions on
curved (3-)spaces, namely the 3-sphere and the 3-hyperboloid. They are
BPS since the time-time component of the metric is constant. 
Then in Sec.~\ref{sec:gravcurvedspace}, we promote the solution on the
3-sphere to be a 
gravitating Skyrmion, for which the Einstein equation fixes the
cosmological constant and the coefficient of the BPS Skyrme term.
Since the second-order equation of motion is harder to solve than the
first-order BPS equation, we propose in Sec.~\ref{sec:specialpot},
a special potential (for which we have no better name than special
potential) which simplifies the Skyrmion equation of motion such that
we in Sec.~\ref{sec:skcurvbg} find analytic solutions on anti-de
Sitter and de-Sitter spacetimes in the limit of vanishing
gravitational coupling (i.e.~the limit of no backreaction onto
gravity). 
These solutions are, however, non-BPS. 
Finally, in Sec.~\ref{sec:gravsk} we take into account (a finite)
coupling of the Skyrmion to gravity for which we have not been able to
find analytic solutions and thus have turned to numerical
calculations. The numerical solutions show that the analytic solutions
are very good approximations for small values of the gravitational
coupling.
The results are summarized and discussed in Sec.~\ref{sec:discussion}.

\section{The model}\label{sec:model}

The model we are interested in in this paper is based on taking the
BPS Skyrme model \cite{Adam:2010fg,Adam:2010ds} and putting it on a
curved background geometry with metric $g_{\mu\nu}$, 
\beq
S &=& \int d^4x \; \sqrt{-g}\,
  \left[\mathcal{L}_{06} + \frac{1}{4\alpha}
  (\mathcal{R}-2\Lambda)\right], \\
\mathcal{L}_{06} &=& \mathcal{L}_6 - V,
\eeq
where $\mathcal{R}$ is the Ricci scalar, $\Lambda$ is the cosmological
constant, $\alpha\equiv 4\pi G$ is the gravitational coupling,
$g=\det g_{\mu\nu}$ and the sextic term, $\mathcal{L}_6$, is given by
the square of the baryon current 
\beq
  \mathcal{L}_6 &=& -4\pi^4c_6
  g_{\mu\nu}\mathcal{B}^\mu\mathcal{B}^\nu, \label{eq:L6}\\
  \mathcal{B}^\mu &\equiv&
  \frac{1}{24\pi^2\sqrt{-g}}\epsilon^{\mu\nu\rho\sigma}\Tr[L_\nu L_\rho L_\sigma],
\eeq
where the left-current, which is $\mathfrak{su}(2)$ valued, is defined
in terms of the derivative of the nonlinear sigma-model field $U$, as 
\beq
L_\mu \equiv U^\dag\p_\mu U,
\eeq
and $U$ is given by $U=\sigma\mathbf{1}_2+i\pi^a\tau^a$,
where $a=1,2,3$ is summed over and $\tau^a$ are the Pauli matrices. 
$V$ is an appropriately chosen potential, which for concreteness we
will choose to be in the form
\beq
V_n &=& \frac{c_{0n}}{n}\left(1 - \frac{1}{2^n}\Tr[U]^n\right),
\label{eq:pot}
\eeq
with $n=1,2$.
For $n=1$ it is simply the standard pion mass term, while for $n=2$ it
is the so-called modified mass
term \cite{Piette:1997ce,Kudryavtsev:1999zm,Kopeliovich:2005vg}, see
also
e.g.~Refs.~\cite{Nitta:2012wi,Gudnason:2013qba,Gudnason:2014gla,Gudnason:2014hsa}. 

The topological charge, which is called the baryon number or Skyrmion
number, is defined as
\beq
B = \int d^3x \; \sqrt{-g} \, \mathcal{B}^0.
\eeq
The charge is independent of the background geometry by construction,
unless a black hole horizon has formed. In that case, the integral is
evaluated only from the horizon to spatial infinity, yielding a
total charge smaller than unity.
If furthermore a cosmological horizon is present, the integral is then
only evaluated up to said horizon, which again may give a total baryon
charge less than one. 

As we want to couple the Skyrmion to gravitational backgrounds, we
need the energy-momentum tensor, which can be written as
\beq
T_{\mu\nu} =
8\pi^4 c_6 \mathcal{B}_\mu \mathcal{B}_\nu
- g_{\mu\nu} \left(4\pi^4 c_6 g_{\rho\sigma}
\mathcal{B}^\rho \mathcal{B}^\sigma -V\right), 
\eeq
which is the energy-momentum tensor of a perfect
fluid \cite{Adam:2010ds,Adam:2014nba,Adam:2014dqa}. 
The Einstein equation reads
\beq
G_{\mu\nu} = 2\alpha T_{\mu\nu} - \Lambda g_{\mu\nu},
\eeq
where the Einstein tensor depends on the background metric.

Let us now consider the Bogomol'nyi completion -- keeping track of the
metric factors -- for the BPS Skyrme model
\beq
\mathcal{L}_{06} =
-g_{\mu\nu} W^\mu W^\nu
\mp 4\pi^2\sqrt{c_6} \left(g_{00}\right)^{-\frac{1}{2}} \sqrt{V}
g_{\mu\nu} \mathcal{B}^\mu \delta^{\nu 0},
\eeq
where we have defined
\beq
W^\mu \equiv 2\pi^2 \sqrt{c_6} \mathcal{B}^\mu
\mp \frac{\delta^{\mu 0}\sqrt{V}}{\sqrt{g_{00}}}.
\eeq
This Bogomol'nyi completion is written in the rest frame of the
Skyrmion (hence the $\delta^{\mu 0}$).
The BPS equation is then simply $W^\mu = 0$, which can be fleshed out 
as
\beq
\frac{\sqrt{c_6}}{12}\epsilon^{\mu\nu\rho\sigma}\Tr[L_\nu L_\rho L_\sigma]
= \pm\frac{\delta^{\mu 0}\sqrt{-g}\sqrt{V}}{\sqrt{g_{00}}},
\label{eq:BPS} 
\eeq
where the upper (lower) signs are for the Skyrmion (anti-Skyrmion)
solution. 
The energy for configurations that saturate the BPS equation is
then the 3-dimensional integral of the cross term coming from the
Bogomol'nyi completion 
\beq
M_{\rm BPS} = \pm\frac{\sqrt{c_6}}{6}\int d^3x\;
\sqrt{g_{00}}\sqrt{V}\epsilon^{ijk} \Tr[L_i L_j L_k]. 
\eeq
Iff the time-time component of the metric is constant (i.e.~not
depending on any spatial coordinate), then the energy is identical to
flat space version of the BPS
Skyrmion \cite{Adam:2010fg,Adam:2010ds}.
Otherwise the static energy experiences a warp factor from the
time-time component of the metric which thus weighs the integral of
the topological charge. Hence the static energy is no longer linearly
proportional to the topological charge, i.e.~the baryon charge. 
This often happens for gravitational versions of solitons, see
e.g.~\cite{Sutcliffe:2011sr,Aldrovandi:2007bn,Bolognesi:2010nb,Elliot-Ripley:2015hoa}. 
Note that the mass is positive (semi-)definite independent of the
sign. 

Using now the BPS equation \eqref{eq:BPS}, the stress-energy tensor
simplifies for BPS saturated configurations to
\beq
T_{\mu\nu} = \frac{2g_{\mu 0}g_{\nu 0}}{g_{00}} \, V. \label{eq:TmunuBPS}
\eeq
From this tensor we can first of all observe that the BPS equation
yields vanishing pressure ($T_{ii}=0$), which was already pointed out
in the flat-space version of this type of BPS
models \cite{Bazeia:2007df}, see
also \cite{Adam:2010ds,Adam:2014nba,Adam:2014dqa}. 
The second observation is that the energy density for BPS saturated
configurations take the form,
\beq
\mathcal{E}_{\rm BPS} = 2V, \label{eq:BPSenergy_density}
\eeq
and hence depend only on the shape of the potential.
This will be important later.

In this paper we will consider only the $B=1$ sector, for which --
with the present type of potential that only depends on $\Tr[U]$ with
the vacuum at $U=\mathbf{1}_2$ -- we
can employ a spherically symmetric Ansatz for the field $U$ as
\beq
U = \mathbf{1}_2\cos f + i\hat{x}^a\tau^a \sin f, \label{eq:hedgehog} 
\eeq
where $\hat{x}^a$ is the Cartesian unit vector and $f=f(r)$ is a
radial function, which we will denote the Skyrmion profile function.

Throughout the paper we will assume a diagonal metric of the type
\beq
ds^2 = N^2 C dt^2 - C^{-1} dr^2
- \Omega\left(d\theta^2 + \sin^2\theta d\phi^2\right),
\label{eq:genericmetric} 
\eeq
with $N=N(r)$, $C=C(r)$ and $\Omega=\Omega(r)$ being radial functions; 
for which the Lagrangian density reads
\beq
\mathcal{L}_6 = -\frac{c_6C\sin^4(f) f_r^2}{\Omega^2},
\eeq
where $f_r\equiv\p_rf$ and we have used the
Ansatz \eqref{eq:hedgehog}.
The BPS equation \eqref{eq:BPS}, with these Ans\"atze reads
\beq
\sqrt{c_6} \sin^2(f) f_r = \mp\frac{\Omega\sqrt{V}}{\sqrt{C}},
\label{eq:BPSf}
\eeq
which we formally can solve as
\beq
\int df\; \frac{\sin^2f}{\sqrt{V}} =
\mp \frac{1}{\sqrt{c_6}}\int dr\; \frac{\Omega}{\sqrt{C}} + \kappa,
\label{eq:implicitBPSsol}
\eeq
where $\kappa$ is an appropriately chosen constant. 
This implicit solution depends both on the choice of the potential and
the gravitational background.

The baryon-charge density now reads
\beq
\sqrt{-g}\mathcal{B}^0 = -\frac{1}{2\pi^2} \sin\theta\sin^2(f) f_r. 
\eeq
For convenience, we will define the following radial baryon-charge
density
\beq
\mathcal{B}^r\equiv \frac{1}{\Omega}\int d\theta d\phi\;
\sqrt{-g}\mathcal{B}^0
= -\frac{2\sin^2(f) f_r}{\pi\Omega}
= \pm\frac{2}{\pi}\sqrt{\frac{V[f]}{c_6 C}}, \label{eq:Br}
\eeq
where we have integrated the two angular variables and in the last
equality we have used the BPS equation \eqref{eq:BPSf}. 
The baryon number is thus simply $B=\int dr\;\Omega\mathcal{B}^r$. 
In the following, we will use the terms radial baryon-charge density and
baryon-charge density interchangeably. 

The integrated baryon charge can be evaluated as
\beq
B = \int_{r_0}^L dr\; \Omega\mathcal{B}^r
= \frac{2f_0 - \sin 2f_0}{2\pi},
\label{eq:integratedB}
\eeq
where $f_0\equiv f(r_0)$ is the value of the Skyrmion profile function
at $r_0$, which is the radius from where the integral begins and $L$
is the compacton size (size of the compact Skyrmion solution). If a
black hole horizon has formed on the background under consideration,
$r_0=r_h$ is the horizon radius, otherwise $r_0=0$.
In the above expression, we have assumed that $f(L)=0$, which is also
the definition of the compacton size. 

The nonzero components of the stress-energy tensor with the
Ansatz \eqref{eq:hedgehog} and the metric \eqref{eq:genericmetric} are
\beq
T_t^{\phantom{t}t} &=& \frac{c_6 C\sin^4(f) f_r^2}{\Omega^2} + V,
\label{eq:Ttt} \\
T_r^{\phantom{r}r} &=& T_\theta^{\phantom{\theta}\theta}
= T_\phi^{\phantom{\phi}\phi}
= -\frac{c_6C\sin^4(f) f_r^2}{\Omega^2} + V.
\eeq
It is straightforward to check that when the BPS
equation \eqref{eq:BPSf} is satisfied then the system possesses a
vanishing pressure, as shown generally in Eq.~\eqref{eq:TmunuBPS}.
Using again the BPS equation \eqref{eq:BPSf} and the definition of the
radial baryon-charge density \eqref{eq:Br}, we can write the
stress-energy tensor for BPS configurations as
\beq
T_t^{\phantom{t}t} = \frac{c_6\pi^2 C}{2}
  \left(\mathcal{B}^r\right)^2, \qquad
T_r^{\phantom{r}r} = T_\theta^{\phantom{\theta}\theta}
= T_\phi^{\phantom{\phi}\phi}
= 0.
\eeq

Let us first consider the two potentials of Eq.~\eqref{eq:pot}, for
which we can integrate the left-hand side of the BPS solution
\beq
f &=& 2 \arccos\sqrt[3]{\pm\frac{3}{4}\sqrt{\frac{c_{01}}{2c_6}}
  \int_{r_0}^r dr'\; \frac{\Omega}{\sqrt{C}}
  + \cos^3\left(\frac{f_0}{2}\right)},
  \label{eq:fsol1}\\
f &=& \arccos\left[\pm\sqrt{\frac{c_{02}}{2c_6}}
  \int_{r_0}^r dr'\; \frac{\Omega}{\sqrt{C}} + \cos f_0\right]. \label{eq:fsol2}
\eeq
The (radial) baryon-charge densities for these two solutions read
\beq
\mathcal{B}^r &=& \pm\frac{4}{\pi}\sqrt{\frac{c_{01}}{2c_6}}
\sqrt{1-\left(\pm\frac{3}{4}\sqrt{\frac{c_{01}}{2c_6}}
  \int_{r_0}^r dr'\; \frac{\Omega}{\sqrt{C}}
  + \cos^3\left(\frac{f_0}{2}\right)\right)^{\frac{2}{3}}}
  \frac{1}{\sqrt{C}},\\
\mathcal{B}^r &=& \pm\frac{2}{\pi}\sqrt{\frac{c_{02}}{2c_6}}
\sqrt{1-\left(\pm\sqrt{\frac{c_{02}}{2c_6}}
  \int_{r_0}^r dr'\; \frac{\Omega}{\sqrt{C}} + \cos f_0\right)^2}
  \frac{1}{\sqrt{C}}.
\eeq

For completeness, we will review the known analytic solutions for the
BPS Skyrmion in flat space.
In this case, the metric simply reads
\beq
ds^2 = dt^2 - dr^2
- r^2 \left(d\theta^2 + \sin^2\theta d\phi^2\right),
\eeq
which means that $N=C=1$ and $\Omega=r^2$ and therefore the BPS
solutions for the two potentials are given by Eqs.~\eqref{eq:fsol1}
and \eqref{eq:fsol2} with, \cite{Adam:2010fg,Adam:2010ds} 
\beq
f &=& 2 \arccos\sqrt[6]{\frac{c_{01}}{2^5c_6}}r,
  \label{eq:fsol1flat}\\
f &=& \arccos\left[\frac{1}{3}\sqrt{\frac{c_{02}}{2c_6}} r^3 - 1\right],
  \label{eq:fsol2flat}
\eeq
where we have fixed the sign and $f_0$ to match the boundary
conditions of the Skyrmion, as opposed to the anti-Skyrmion. 
Since we are in flat space and thus there is no black hole horizon, we
have fixed the boundary conditions of the solution in the usual way,
such that $f(0)=\pi$ and the size of the Skyrmion -- which is a
compacton -- is given by $f(L)=0$ and it reads, respectively, for the
two solutions
\beq
L = \sqrt[6]{\frac{2^5c_6}{c_{01}}}, \qquad
L = \sqrt[6]{\frac{2^3 3^2c_6}{c_{02}}}.
\eeq
The baryon-charge densities for the two solutions read
\beq
\mathcal{B}^r &=& \frac{4}{\pi}\sqrt{\frac{c_{01}}{2c_6}}
\sqrt{1-\sqrt[3]{\frac{c_{01}}{2^5 c_6}}r^2}, \\
\mathcal{B}^r &=& \frac{2}{\pi}
\left(\frac{c_{02}}{2c_6}\right)^{\frac{3}{4}}
\frac{r^{\frac{3}{2}}}{\sqrt{3}}
\sqrt{2-\frac{r^3}{3\sqrt{2}}}.
\eeq

Finally, we should check under what circumstances the BPS equation
solves the full equation of motion, which reads
\beq
2c_6\sin^2f\p_r\left(\frac{NC}{\Omega}\sin^2(f)f_r\right)
-N\Omega\frac{\p V}{\p f} = 0. \label{eq:eomf}
\eeq
By inserting the BPS equation \eqref{eq:BPSf} two times into the above
equation of motion, it reduces to
\beq
\sqrt{c_6}\sin^2f\,\p_r\big(\sqrt{N^2C}\big)\sqrt{V} = 0,
\label{eq:curvedBPScondition}
\eeq
which means that the BPS equation \emph{only} solves the full
equations of motion when the time-time component of the metric is
constant.
This implies that the Skyrmion in the BPS Skyrme model can be BPS on a
curved space, but not on a curved spacetime.
This is also related to the fact that the static energy of the
Skyrmion is only topological when the time-time component of the
metric is constant.

As already mentioned, the BPS condition is equivalent to requiring
that the pressure vanishes ($T_i^{\phantom{i}i}=0$). One may ask
whether considering the non-BPS extension with constant
pressure \cite{Adam:2014nba,Adam:2015lpa} may
ameliorate the problem of the BPS equation not solving the
second-order equation of motion on curved spacetime backgrounds. It
turns out that one simply gets the same condition as
Eq.~\eqref{eq:curvedBPScondition}, with $V\to V-P$, where $P$ is a
constant pressure and so even the corresponding ``non-BPS'' equation
(i.e.~the BPS equation with $V\to V-P$) still requires
$\p_r(\sqrt{N^2C})=0$.
This is in fact expected, since the constant pressure is only the
conservation of the stress-energy tensor of a static Skyrmion on flat
space (Minkowski space), whereas on a curved spacetime background, the
conservation of the stress-energy tensor is $\nabla_\mu T^{\mu\nu}=0$,
with $\nabla_\mu$ being the covariant derivative, e.g.~in the radial
direction, we have
\beq
\nabla_\mu T_r^{\phantom{r}\mu} =
\p_\mu T_r^{\phantom{r}\mu}
-T_t^{\phantom{t}t}\left(\frac{C'}{2C} + \frac{N'}{N}\right)
+T_r^{\phantom{r}r}\left(\frac{C'}{2C} + \frac{N'}{N}
  + \frac{\Omega'}{\Omega}\right)
-\left(T_\theta^{\phantom{\theta}\theta} +
T_\phi^{\phantom{\phi}\phi}\right)\frac{\Omega'}{2\Omega} = 0, \nonumber
\eeq
where we have assumed a diagonal stress-energy tensor.
The extra terms involving Christoffel symbols will in general not give 
rise to a constant (isotropic) pressure (but $C=N={\rm const}$ will).

\section{BPS solutions}

\subsection{The BPS Skyrmion on curved space}\label{sec:curvedspace}

In this section we will put the BPS Skyrme model on a curved 3-space,
which is not a consistent gravitational background (in the absence of
matter). The spatial part of the manifold is conformally flat and thus
the spaces we consider here correspond to $\mathbb{R}\times S^3$ and 
$\mathbb{R}\times\mathbb{H}^3$. These spaces were considered for the
normal Skyrme model in
Refs.~\cite{Manton:1986pz,Manton:1990gr,Atiyah:2004nh,Winyard:2015ula}. 
The metric can be written as
\beq
ds^2 = dt^2 - \frac{dr^2}{1-\Lambda r^2} 
- r^2\left(d\theta^2 + \sin^2\theta d\phi^2\right).
\label{eq:curved_space_metric}
\eeq
Comparing with the metric \eqref{eq:genericmetric}, we have
$N^2C=1$, $C=1-\Lambda r^2$ and $\Omega=r^2$. This means that the
solution to the BPS equation is also a solution to the second-order
equation of motion. Plugging these functions for $C$ and $\Omega$ into
the solutions \eqref{eq:fsol1} and \eqref{eq:fsol2}, we get
\begin{align}
f &= 2 \arccos\sqrt[3]{\frac{3}{8}\sqrt{\frac{c_{01}}{2c_6}}
  \left(\frac{\arcsin\sqrt{\Lambda}r}{\Lambda^{\frac{3}{2}}}
  - \frac{r\sqrt{1-\Lambda r^2}}{\Lambda}\right)},
  \label{eq:fsol1scbg}\\
f &= \arccos\left[\frac{1}{2}\sqrt{\frac{c_{02}}{2c_6}}
  \left(\frac{\arcsin\sqrt{\Lambda}r}{\Lambda^{\frac{3}{2}}}
  - \frac{r\sqrt{1-\Lambda r^2}}{\Lambda}\right) - 1\right],
  \label{eq:fsol2scbg}
\end{align}
for the two potentials under consideration,
respectively.
The above solutions are valid for both $\Lambda>0$ and $\Lambda<0$;
in the limit $\Lambda\to 0$, they are equal to the flat-space ones of
Eqs.~\eqref{eq:fsol1flat} and \eqref{eq:fsol2flat}.

The solutions are only well defined for $r\leq 1/\sqrt{|\Lambda|}$
(for $S^3$ this is also the allowed range of the coordinate itself).
In order for the compacton to contain a full unit of baryon charge, we
get the following constraints
\beq
\sqrt[6]{\frac{c_{01}}{2c_6}}
\sqrt[3]{\frac{3\pi}{16}}\frac{1}{\sqrt{|\Lambda}|} \geq 1, \qquad
\frac{\pi}{8}
\sqrt{\frac{c_{02}}{2c_6}}\frac{1}{|\Lambda|^{\frac{3}{2}}} \geq 1,
\eeq
for the two potentials, respectively.

The baryon-charge densities for the above solutions read
\beq
\mathcal{B}^r &=& \frac{4}{\pi}\sqrt{\frac{c_{01}}{2c_6}}
\sqrt{1-\left[\frac{3}{8}\sqrt{\frac{c_{01}}{2c_6}}\left(
  \frac{\arcsin\sqrt{\Lambda}r}{\Lambda^{\frac{3}{2}}}
  -\frac{r\sqrt{1-\Lambda r^2}}{\Lambda}
  \right)\right]^{\frac{2}{3}}}\frac{1}{\sqrt{1-\Lambda r^2}}, \\ 
\mathcal{B}^r &=& \frac{2}{\pi}\sqrt{\frac{c_{02}}{2c_6}}
\sqrt{1-\left[\frac{1}{2}\sqrt{\frac{c_{02}}{2c_6}}
  \left(\frac{\arcsin\sqrt{\Lambda}r}{\Lambda^{\frac{3}{2}}}
  - \frac{r\sqrt{1-\Lambda r^2}}{\Lambda}\right) - 1\right]^2}
  \frac{1}{\sqrt{1-\Lambda r^2}}.
\eeq

\begin{figure}[!t]
\begin{center}
\mbox{
\subfloat[]{\includegraphics[width=0.49\linewidth]{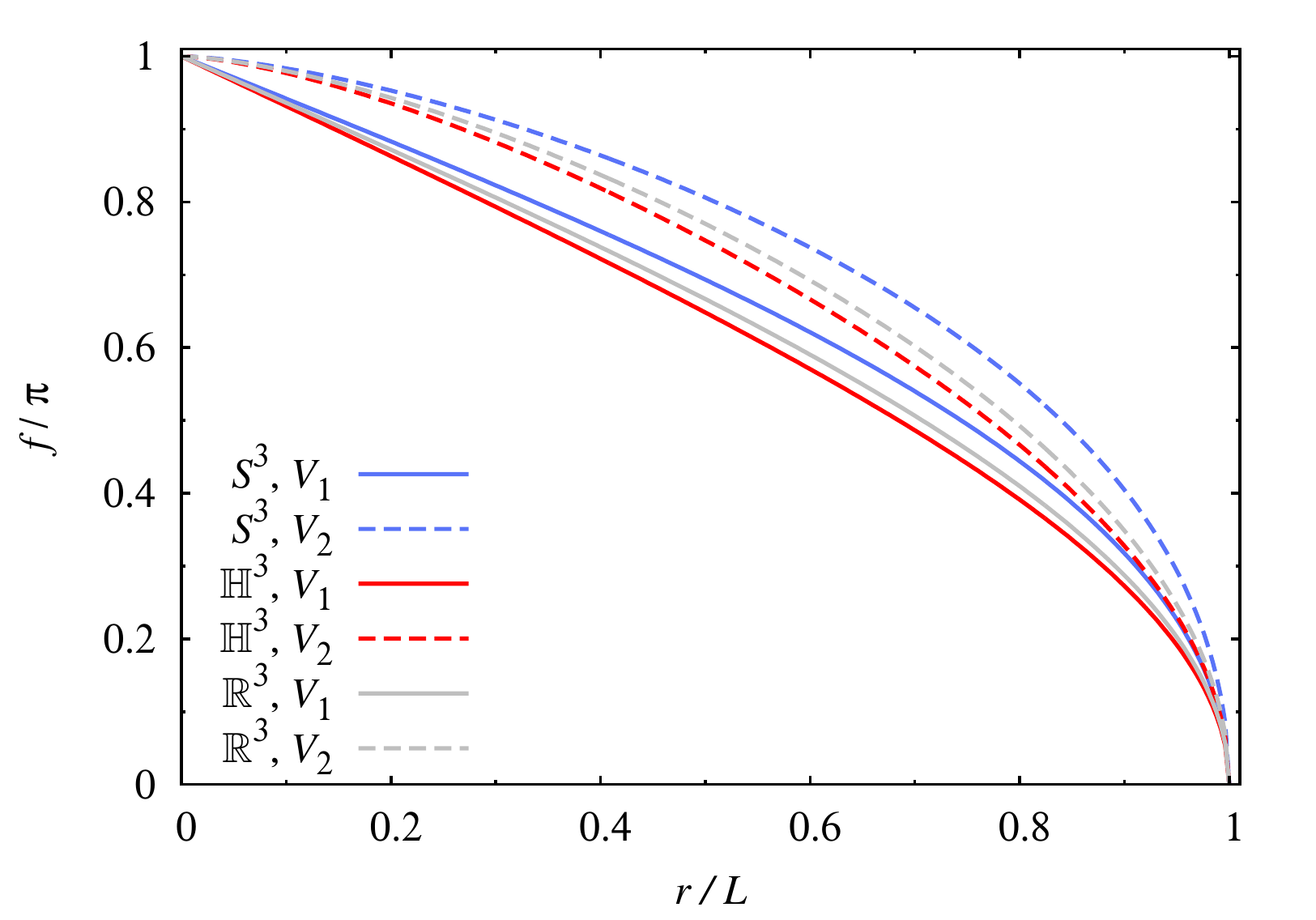}}\
\subfloat[]{\includegraphics[width=0.49\linewidth]{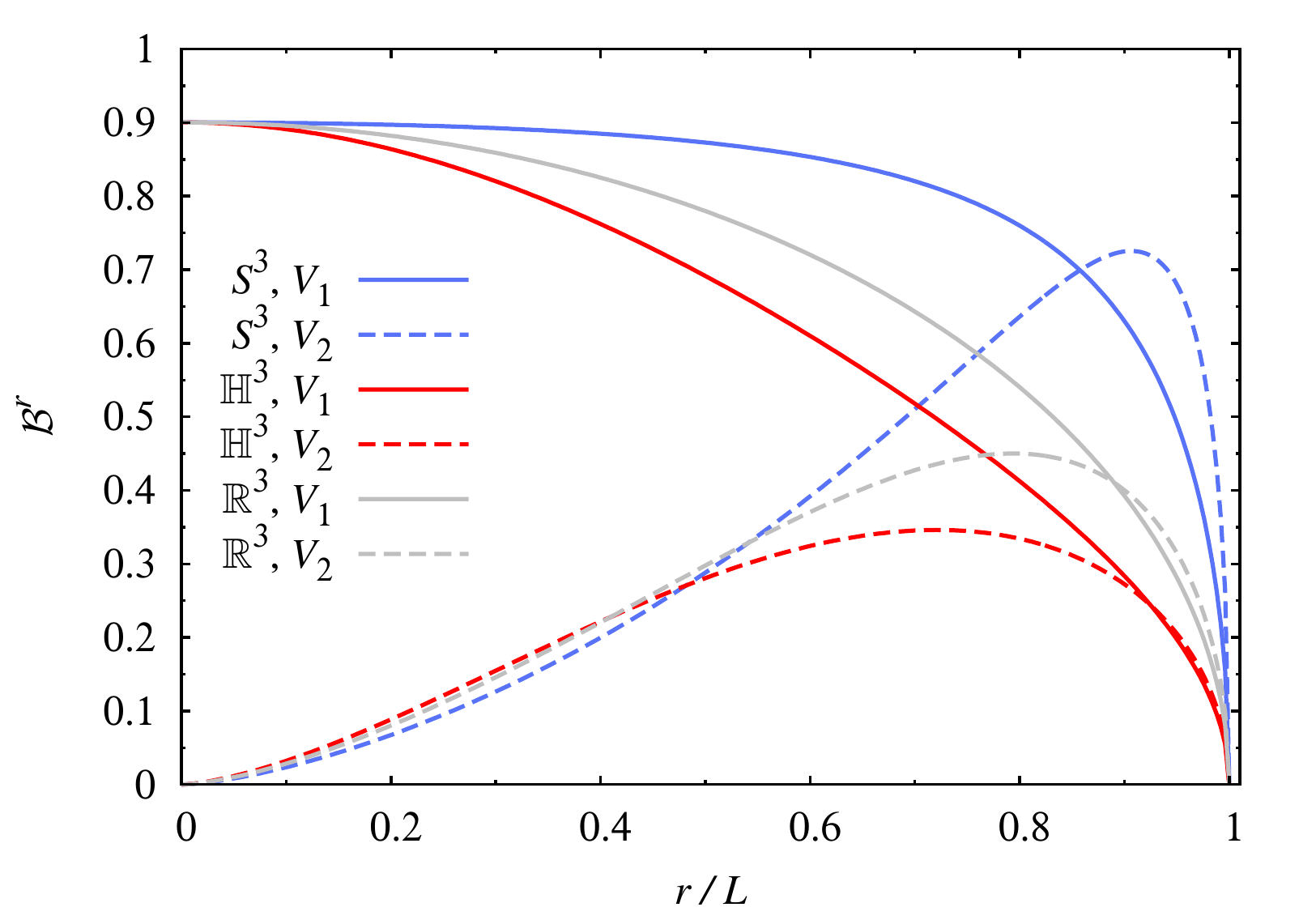}}}
\mbox{
\subfloat[]{\includegraphics[width=0.49\linewidth]{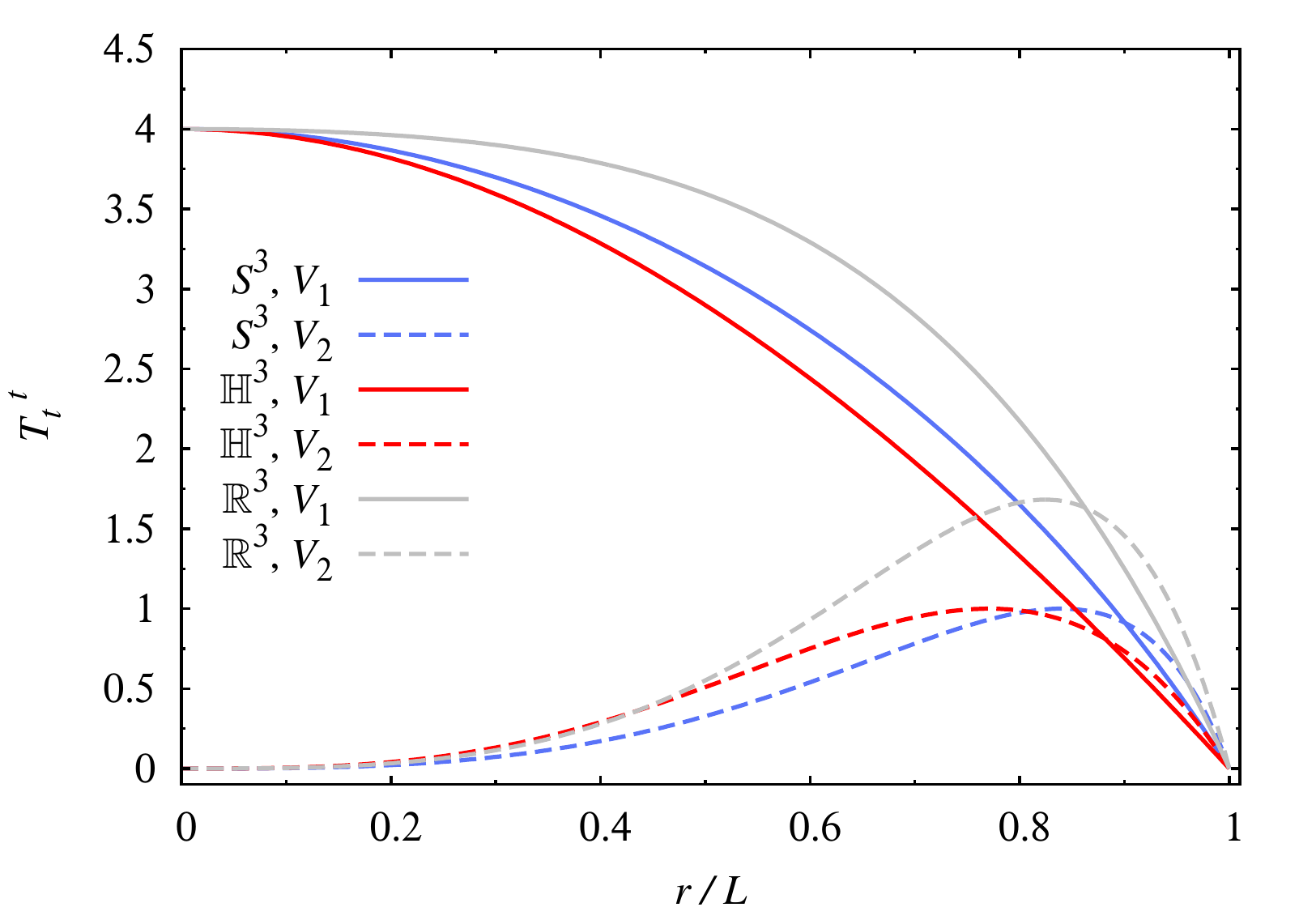}}}
\caption{(a) Profile functions, $f$ (b) baryon-charge densities
  and (c) energy densities for analytic Skyrmion solutions on curved
  spatial backgrounds: $S^3, \mathbb{H}^3$ with $\Lambda=\pm 1/4$ for
  two choices of potentials: $V_{1,2}$. 
  For comparison, the flat-space solution ($\mathbb{R}^3$) is shown as
  well.
  For this figure $c_{01}=c_{02}=c_6=1$.
  In the figure the solutions are rescaled by their respective compacton
  size: $L_{S^3,V_1}=1.62202$, $L_{S^3,V_2}=1.78883$,
  $L_{\mathbb{H}^3,V_1}=1.91009$, $_{\mathbb{H}^3,V_2}=2.22702$,
  $L_{\mathbb{R}^3,V_1}=2^{\frac{5}{6}}$ and
  $L_{\mathbb{R}^3,V_2}=3^{\frac{1}{3}}\sqrt{2}$. 
}
\label{fig:scbg}
\end{center}
\end{figure}

In Fig.~\ref{fig:scbg} are shown plots of the profile functions, the 
baryon-charge densities and the energy densities for the two solutions
(corresponding to the two different potentials, $V_{1,2}$) for
$\Lambda=\pm 1/4$.
For reference, we have shown the flat-space
solutions \eqref{eq:fsol1flat} and \eqref{eq:fsol2flat}.
As can readily be seen from Fig.~\ref{fig:scbg}b, the baryon-charge
density vanishes at $r=0$ for $V=V_2$. This follows from
Eq.~\eqref{eq:Br} because the right-hand side is proportional to
$\sqrt{V}$ and hence for BPS solutions, when the potential vanishes,
so does the baryon-charge density.

The solutions \eqref{eq:fsol1scbg} and \eqref{eq:fsol2scbg} solve the
BPS equation and the full second-order equation of motion for the
Skyrmion fields as well as the spatial components of the Einstein
equations. This is because the spatial part of the metric is
conformally flat and the BPS equation ensures that there is no
pressure in the system and hence the spatial components of the
stress-energy tensor vanish. The time-time component of the Einstein
equation is not satisfied in this case, both because the background
itself is not a consistent gravitational background (without matter)
and the gravitational backreaction has not been taken into account. 

The time-time component of the Einstein equation can be satisfied by
neglecting the gravitational backreaction, $\alpha=0$, and setting the 
cosmological constant to zero: $\Lambda=0$. This, however, corresponds
exactly to the flat-space solutions in Eqs.~\eqref{eq:fsol1flat}
and \eqref{eq:fsol2flat}.

The time-time component of the Einstein tensor is equal to the scalar
curvature in this case, which is $2\Lambda$. A solution to the
Einstein equation thus requires a constant energy density (constant
$T_0^{\phantom{0}0}$). Although this is possible in the standard Skyrme
model without a potential term \cite{Ayon-Beato:2015eca}, with for
instance the identity map from $S^3$ to $S^3$, the BPS Skyrme model
requires a special potential.
In fact, from Eq.~\eqref{eq:BPSenergy_density} we can clearly read off 
which potential is required for the BPS model to have a constant
energy density; namely a constant nonzero potential.
We will consider this case in the next section.

\subsection{The gravitating BPS Skyrmion on curved space}
\label{sec:gravcurvedspace}

In order for the Skyrmion to be BPS and to solve the second-order
equations of motion, we need $g_{00}$ to be a constant and thus for
space to be isotropic and homogeneous, it must be conformally flat and 
take the form of Eq.~\eqref{eq:curved_space_metric} (although many
other coordinates may be used for the same space).
In the previous section we have put BPS Skyrmions on curved spaces
with the latter metric. These BPS Skyrmions solve the second-order
equations of motion as well as the spatial part of the Einstein
equations.
In order for a BPS Skyrmion to solve all the Einstein equations, we
need to match the scalar curvature --  resulting from the Einstein
tensor -- with the energy density of the Skyrmion. Since the spaces of
Eq.~\eqref{eq:curved_space_metric} have constant scalar curvature, we
need to choose the following potential
\beq
V = c_0,
\eeq
where $c_0>0$ is a positive real constant.
On an infinite space, this would imply a solution with infinite
energy, but on $S^3$, which is compact, this gives rise to a
finite-energy solution.

The BPS solution is thus implicit and reads
\beq
\frac{1}{4}\left(2f-\sin 2f\right) = \frac{\pi}{2} 
- \frac{1}{2}\sqrt{\frac{c_0}{c_6}}
\left(\frac{\arcsin\sqrt{\Lambda}r}{\Lambda^{\frac{3}{2}}}
  - \frac{r\sqrt{1-\Lambda r^2}}{\Lambda}\right),
  \label{eq:gravitationalBPSsolf}
\eeq
with $\Lambda>0$. 
Since BPS-ness implies vanishing pressure ($T_{ii}=0$) and the
background solves the spatial parts of the Einstein equations, only 
the time-time component of the Einstein equation remains, which is
solved by
\beq
2\Lambda = 2\alpha \mathcal{E},
\eeq
where the energy density of the compacton is
\beq
\mathcal{E} = \left\{
\begin{array}{ll}
2c_0, & 0 \leq r \leq L\\
c_0, & r > L
\end{array}\right.,
\eeq
which means that we can only solve the time-time component of the
Einstein equation by setting the size of the compacton as
$L=\frac{1}{\sqrt{\Lambda}}$. In other words, we need to cover the
whole 3-sphere for the solution to solve the Einstein equations.
This is also natural.\footnote{Actually a more contrived solution can
be constructed by making a step potential
\beq
V = c_0\left[1+\Theta(r-L)\right],\nonumber
\eeq
where $\Theta$ is the Heaviside step function. This potential
compensates the missing energy density from the baryon-charge density
(squared) so that the energy density is constantly equal to $2c_0$
over all the 3-sphere but the size of the compacton is
$L\leq 1/\sqrt{\Lambda}$. 
} 
The solution hence determines the cosmological constant as
\beq
\Lambda = 2\alpha c_0. \label{eq:Lambdasol}
\eeq
This solution is consistent on $\mathbb{R}\times S^3$ and indeed a
full gravitational BPS Skyrmion.

The mentioned constraint on the compacton size, $L=1/\sqrt{\Lambda}$,
which means that $f(L)=0$, translates into the following constraint
\beq
\frac{c_0}{c_6} = 4\Lambda^3.
\eeq
Combining this with Eq.~\eqref{eq:Lambdasol} determines $c_6$ as
\beq
c_6 = \frac{1}{8\alpha\Lambda^2} = \frac{1}{32\alpha^3 c_0^2}.
\eeq

The baryon-charge density for this fully gravitating BPS Skyrmion
reads
\beq
\mathcal{B}^r = \frac{4\Lambda^{\frac{3}{2}}}{\pi}
\frac{1}{\sqrt{1-\Lambda r^2}}.
\eeq
As can be seen from the above expression, the baryon-charge density
blows up at $r=L=1/\sqrt{\Lambda}$, but the total baryon charge
\beq
B = \frac{4\Lambda^{\frac{3}{2}}}{\pi}\int_0^{1/\Lambda^2} dr\;
\frac{r^2}{\sqrt{1-\Lambda r^2}} = 1,
\eeq
is finite and indeed integrates to unity as it should.

\begin{figure}[!t]
\begin{center}
\mbox{
\subfloat[]{\includegraphics[width=0.49\linewidth]{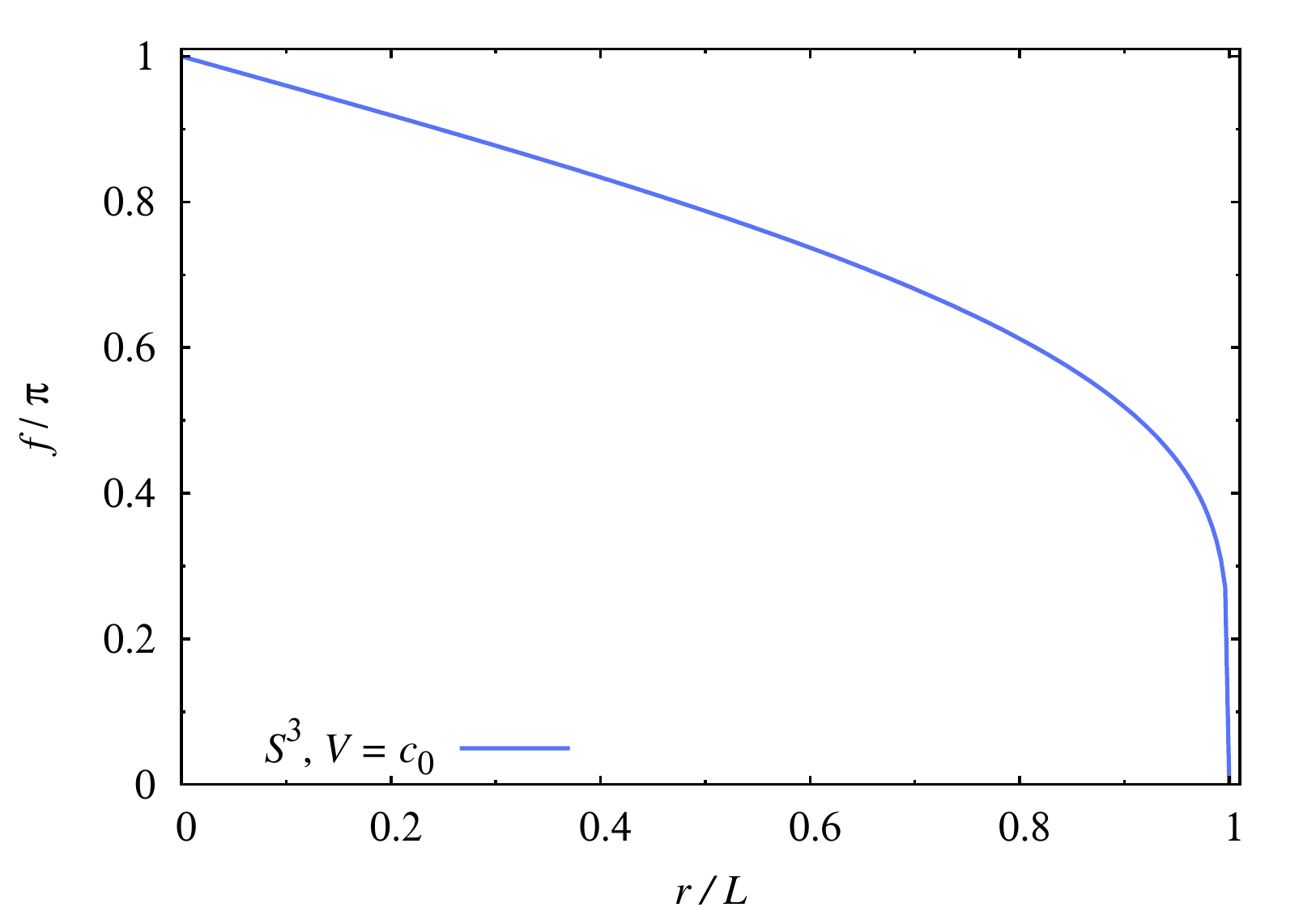}}\
\subfloat[]{\includegraphics[width=0.49\linewidth]{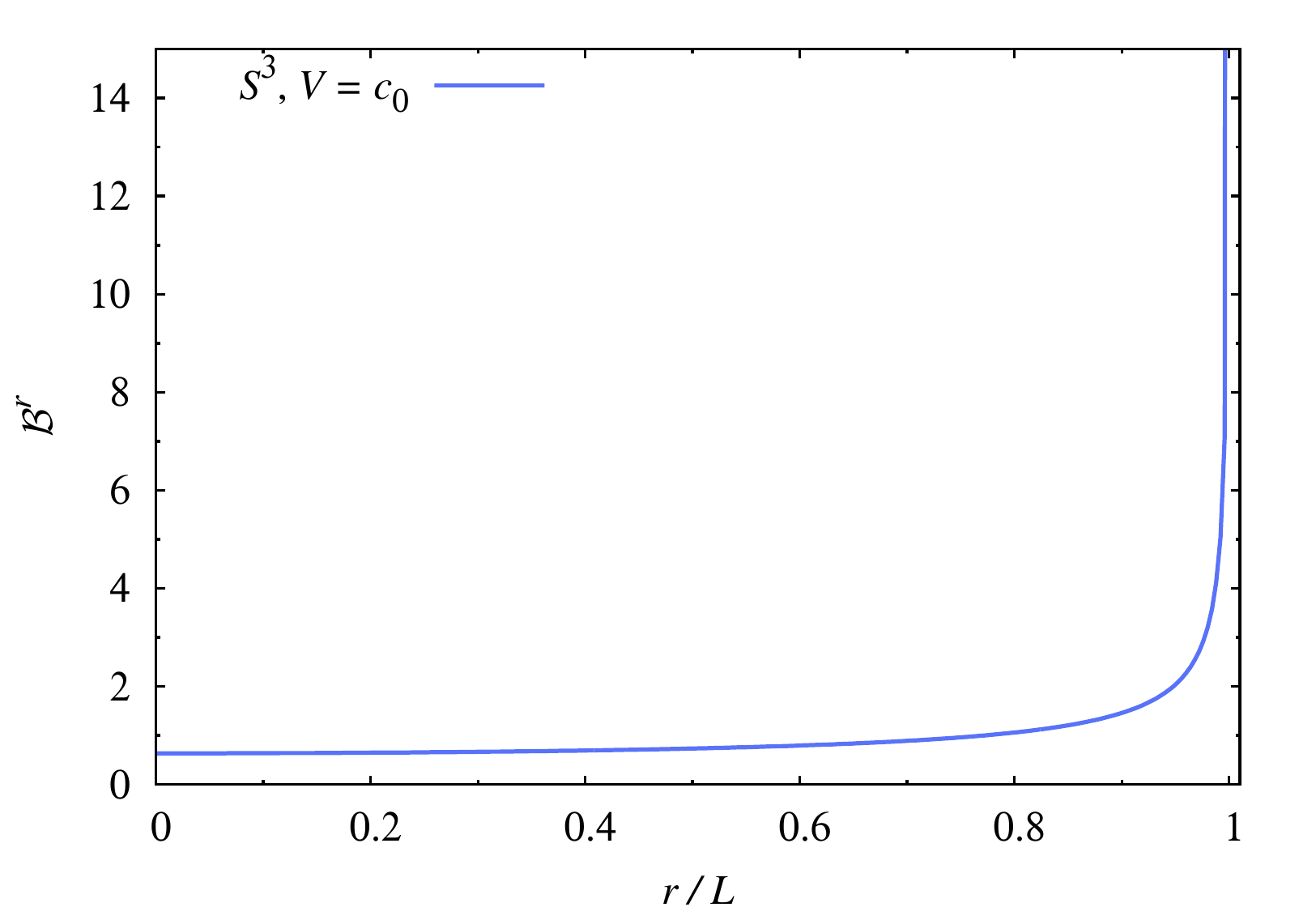}}}
\caption{(a) Profile function, $f$, and (b) baryon-charge density
  for the analytic gravitating BPS Skyrmion on the curved spatial
  background, $\mathbb{R}\times S^3$ for $\Lambda=1/4$ with a constant
  potential. 
  For this figure $c_{0}=c_6=1$.
  In the figure the solution is rescaled by its respective compacton
  size: $L_{S^3,V=c_0}=1/\sqrt{\Lambda}=2$. 
}
\label{fig:scbg2}
\end{center}
\end{figure}

In Fig.~\ref{fig:scbg2} are shown plots of the profile function and
baryon-charge density for the fully gravitating BPS solution on
$\mathbb{R}\times S^3$ for $\Lambda=1/4$. We do not show the energy
density as it is simply given by the constant
$T_t^{\phantom{t}t}=2c_0$.

Although the time-time component of the Einstein equations allows for
a hyperbolic space by considering a negative $c_0<0$, this does not
give a real-valued solution for $f$ from
Eq.~\eqref{eq:gravitationalBPSsolf}. 

The solution \eqref{eq:gravitationalBPSsolf} is similar to that of
Ref.~\cite{Ayon-Beato:2015eca} for the normal Skyrme model.
The latter solution fixes the Skyrme-term coefficient and the radius
of the 3-sphere in terms of the gravitational coupling, the
coefficient of the kinetic term and the cosmological constant.
In our solution, on the other hand, we fix the BPS-Skyrme term
coefficient and the cosmological constant in terms of the
gravitational coupling and the potential.

A comment in store is about the baryon-charge density. Normally, the 
baryon-charge density vanishes at the compacton radius since it is
given by Eq.~\eqref{eq:Br}; from the middle equality it is a product
of $\sin^2f$ and $f_r$ (over $\Omega=r^2$) and hence when $f=0$, it is 
natural that the baryon-charge density also vanishes. In the above
solution, the derivative of $f$ diverges at $r=1/\sqrt{\Lambda}$. 
The compacton fills the 3-sphere and
the choice of coordinates on the 3-sphere yields a baryon-charge
density that diverges at $r=L=1/\sqrt{\Lambda}$. The integral of the 
baryon-charge density -- the total baryon charge -- is however
finite and equals one.
All physical observables are thus continuous and convergent in the
solution. 
The mentioned divergence in the derivative is due to the choice of
coordinates on the 3-sphere. By switching to, perhaps more natural
coordinates
\beq
ds^2 = dt^2 - r_0^2\left(d\psi^2 + \sin^2\psi d\theta^2
+ \sin^2\psi\sin^2\theta d\phi^2\right),
\label{eq:3sphere_hypercoordinates}
\eeq
we can show that the solution is regular.
These coordinates are called hyperspherical coordinates and the angles
take the values $\psi\in[0,\pi]$, $\theta\in[0,\pi]$ and
$\phi\in[0,2\pi)$. The radius of the 3-sphere is now given by $r_0$.
With these coordinates, the appropriate Ansatz for the Skyrme field is
($\psi\to -\psi$ gives the corresponding anti-Skyrmion)
\beq
U = \mathbf{1}_2\cos\psi
-i\tau^1\sin\psi\cos\theta
-i\tau^2\sin\psi\sin\theta\cos\phi
-i\tau^3\sin\psi\sin\theta\sin\phi.
\label{eq:Uhyperspherical}
\eeq
This is simply the identity map from the 3-sphere to the 3-sphere and
it solves the equations of motion for the matter fields (but not
necessarily the BPS equation). 
The BPS equation now reads
\beq
\frac{\sqrt{c_6}}{r_0^3} = \sqrt{c_0},
\eeq
which we can solve by choosing the radius of the 3-sphere as
\beq
r_0 = \sqrt[6]{\frac{c_6}{c_0}}.
\eeq
The baryon-charge density is now
\beq
\mathcal{B}^0 = \frac{1}{2\pi^2r_0^3},
\eeq
which is the inverse of the volume of the
3-sphere \eqref{eq:3sphere_hypercoordinates}, giving rise to a unit
baryon charge on the 3-sphere 
\beq
B = \int d^3x\;\sqrt{-g}\mathcal{B}^0
= \frac{1}{2\pi^2r_0^3}\int_0^\pi d\psi\int_0^\pi
  d\theta \int_0^{2\pi} d\phi\; r_0^3 \sin^2\psi\sin\theta
= 1.
\eeq
The spatial components of the Einstein equations relate the
cosmological constant to the radius of the 3-sphere as
$\Lambda=1/r_0^2$, and thus
\beq
\Lambda = \sqrt[3]{\frac{c_0}{c_6}}.
\eeq
Finally, the time-time component of the Einstein
equations determine the coefficient of the BPS Skyrme term as
\beq
c_6 = \frac{1}{64\alpha^3 c_0^2}.
\eeq
What we have done now is merely changing coordinates. In these
coordinates on the 3-sphere, however, the Skyrmion is an everywhere 
regular function and the baryon-charge density has no divergences.
Up to factors of two (due to different normalization of the new
coordinates), the solution is of course physically the same as
before.

One may notice that our solution, although similar to that of
Ref.~\cite{Ayon-Beato:2015eca}, has one less free parameter than
theirs. This is due to the BPS condition implying vanishing pressure,
which in turn directly relates the radius of the 3-sphere to the
cosmological constant. In that sense, the BPS-ness imposes an
additional constraint on the solution and thus we have one less free
parameter, compared to the solution of Ref.~\cite{Ayon-Beato:2015eca}. 

A final comment about the constant potential is that the Skyrmion
field $f$ does not have a particular vacuum. The charge-one Skyrmion,
however, is topological and wraps the 3-sphere once. It solves the
equations of motion and the BPS equation. The Bogomol'nyi bound
further guarantees that the solution minimizes the static
energy. Since the field cannot unwrap the 3-sphere it is topologically
protected.

\section{Non-BPS solutions}

\subsection{Special potential}\label{sec:specialpot}

\begin{figure}[!th]
\begin{center}
\includegraphics[width=0.5\linewidth]{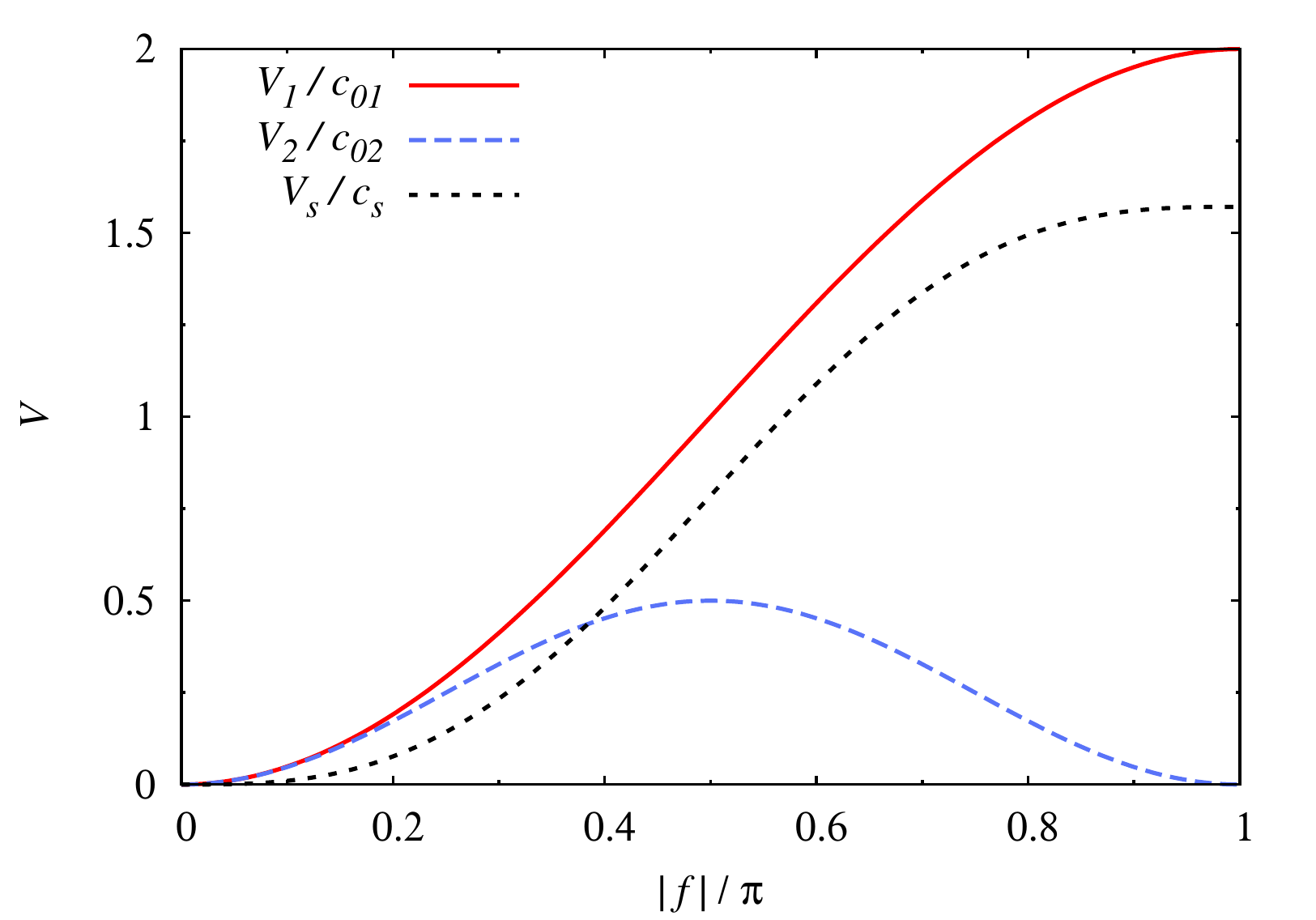}
\caption{The special potential $V_s$ of Eq.~\eqref{eq:specialpot}
alongside with the potentials $V_1$, $V_2$ of Eq.~\eqref{eq:pot}. }
\label{fig:pots}
\end{center}
\end{figure}

Although Eq.~\eqref{eq:curvedBPScondition} throws a monkey wrench in 
using the BPS equation -- which is easier to solve than the
second-order equation of motion -- for gravitational backgrounds with
curved spacetime (i.e.~having also a non-constant $g_{00}$), there
exists a special potential which makes it easier to solve the
second-order matter equation \eqref{eq:eomf}, namely 
\beq
V_s = \frac{c_s}{4}\left|2f - \sin 2f\right|,
\label{eq:specialpot}
\eeq
or in terms of $\Tr[U]$,
\beq
V_s = \frac{c_s}{4}\left|2\arccos\left[\frac{1}{2}\Tr[U]\right]
  - \Tr[U]\left(1 - \frac{1}{4}\Tr[U]^2\right)\right|.
\eeq
The potential is plotted in Fig.~\ref{fig:pots} along with the potential
of Eq.~\eqref{eq:pot} with $n=1,2$.\footnote{Actually this potential
was considered already in the BPS Skyrme
model \cite{Adam:2012hh,Adam:2014nba}. For BPS systems, however, it is
not as special as in our case, because the BPS equation can handle a
very large class of potentials. } 

The reason for this potential being special, is that
\beq
\p_f V_s=c_s\sign(f)\sin^2f,
\eeq
which simplifies the second-order equation of
motion for $f$ to
\beq
2c_6\p_r\left(\frac{NC}{\Omega}\sin^2(f)f_r\right)
-c_s \sign(f) N\Omega = 0, \label{eq:eomfsimplified}
\eeq
which we can integrate right away
\beq
\frac{NC}{\Omega}\sin^2(f)f_r = \frac{c_s}{2c_6} \sign(f)
\left(\int_{r_0}^r dr'\; N\Omega - \kappa_1\right),
\eeq
where $r_0$ is the radius from where the integral begins.
If a black hole horizon has formed on the background under
consideration, $r_0=r_h$ is the horizon radius, otherwise $r_0=0$.
From now on, we will consider only Skyrmion solutions (as opposed to
anti-Skyrmion solutions) without loss of generality (as they are
related by $f\to 2\pi-f$. Therefore $f$ takes on values in the range 
$f\in[0,\pi]$ and $\sign(f)=1$ in the following. 

The implicit solution in terms of $f$ reads
\beq
\frac{1}{4}\left(2f - \sin 2f\right)
= \frac{1}{4}\left(2f_0 - \sin 2f_0\right)
+ \frac{c_s}{2c_6}\int_{r_0}^r dr'\; \frac{\Omega}{NC}
\left(\int_{r_0}^{r'} dr''\; N\Omega - \kappa_1\right),
\label{eq:specialsolf}
\eeq
where $f_0=f(r_0)$ is the value of the profile function at $r_0$.

An important observation is that unless $\kappa_1>0$, the double
integral is positive semi-definite.
This means that if $f_0=\pi$, there is no way for $f$ to flow to its
vacuum value $f=0$. The same conclusion holds for the anti-Skyrmion.

The next observation is that the outer integral contains a factor of
$C$ in the denominator. Since a black hole horizon has a
multiplicity-one pole in $1/C$, $N=1$ at the horizon and $\Omega=r^2$
in the type of spacetime we are considering here, then unless the
parenthesis in the outer integral vanishes at the horizon, the outer
integral will pick up a logarithmic divergence.
Now since $\int_{r_h}^{r_h} dr''$ vanishes, $\kappa_1$ must vanish
when a black hole horizon is present.
This means, however, that if a black hole horizon is present, no
regular Skyrmion (or anti-Skyrmion) solutions with finite energy
exists\footnote{We can easily make a regular infinite-energy
solution, by letting $f$ flow from $0$ to $\pi$ and having
$\kappa_1=0$.}.
Nevertheless, if no black hole horizon is present, we can still have a
regular Skyrmion solution. 

As it is nontrivial to get the explicit solution for $f$, it may be
useful to notice that in terms of the baryon-charge density, we have 
\beq
\mathcal{B}^r = -\frac{2\sin^2(f)f_r}{\pi\Omega}
= -\frac{c_s}{\pi c_6 N C}
\left(\int_{r_0}^r dr'\; N\Omega - \kappa_1\right).
\label{eq:B0specialsol}
\eeq
It is interesting to note that the baryon-charge density of this
solution is dependent on $f$ only through the backreaction of the
Skyrmion onto gravity. If the gravitational coupling, $\alpha$, is
sent to zero, then the baryon-charge density is entirely determined by
the background geometry of spacetime.

The physical meaning of the integration constant $\kappa_1$ is related
the baryon-charge density at $r_0$, as\footnote{This is also
consistent with the above reasoning claiming that $\kappa_1$ should
vanish when a black hole horizon is present. }
\beq
\kappa_1 = \left.\frac{\pi c_6}{c_s} N C\mathcal{B}^r\right|_{r=r_0}.
\eeq

The two components of the stress-energy tensor corresponding to the
energy density and the pressure are
\begin{align}
T_t^{\phantom{t}t} &= T + V, \qquad
T_r^{\phantom{r}r} = T_\theta^{\phantom{\theta}\theta}
= T_\phi^{\phantom{\phi}\phi} = -T + V,
\label{eq:specialpot_energy_pressure}\\
T &\equiv \frac{c_s^2}{4c_6N^2C}
\left(\int_{r_0}^r dr'\; N\Omega - \kappa_1\right)^2, \\
V &\equiv \frac{c_s}{4}\left(2f_0 - \sin 2f_0\right)
+ \frac{c_s^2}{2c_6}\int_{r_0}^r dr'\; \frac{\Omega}{NC}
\left(\int_{r_0}^{r'} dr''\; N\Omega - \kappa_1\right).
\end{align}
For the solution to be BPS we need the following condition to be true
\begin{align}
\frac{c_6}{4c_s}\left(2f_0 - \sin 2f_0\right)
+\frac{1}{2}\int_{r_0}^r dr'\; \frac{\Omega}{NC}
\left(\int_{r_0}^{r'} dr''\; N\Omega - \kappa_1\right)
= \frac{1}{4N^2 C}
\left(\int_{r_0}^r dr'\; N\Omega - \kappa_1\right)^2,
\end{align}
which generically is not satisfied (and hence not BPS).
This condition is tantamount to requiring the pressure to vanish, see
Eq.~\eqref{eq:specialpot_energy_pressure}.
Taking the derivative with respect to $r$ on both sides of the above
equation gives us the following condition
\beq
\p_r\left(\frac{1}{N^2C}\right)
\left(\int_{r_0}^r dr'\; N\Omega - \kappa_1\right) = 0,
\label{eq:curvedBPScondition2}
\eeq
which is consistent with the condition \eqref{eq:curvedBPScondition},
coming from the equation of motion.

\subsection{The Skyrmion with special potential on curved spacetime}
\label{sec:skcurvbg} 

In this section we start by putting the Skyrmion with the special
potential \eqref{eq:specialpot} on a curved spacetime background.
Since a curved spacetime implies that $N^2C$ is not a constant,
the BPS condition \eqref{eq:curvedBPScondition} is not satisfied and
thus the solutions in this section are not BPS, see also
Eq.~\eqref{eq:curvedBPScondition2}.

In this section we neglect the backreaction of the Skyrmion to
gravity, i.e.~we set the gravitational coupling $\alpha=0$. This limit
is a good approximation when the energy/mass scale of the Skyrmion is
very small compared to that of gravity, i.e.~$1/\sqrt{G}$.

The spacetimes we consider in this section are pure anti-de Sitter and
de Sitter spaces, for which we will choose global coordinates or in
the case of de Sitter, static coordinates
\beq
ds^2 = N^2 C dt^2 - C^{-1}dr^2
- \Omega\left(d\theta^2 + \sin^2\theta d\phi^2\right),
\eeq
where
\beq
N = 1, \qquad
C = 1\pm\frac{r^2}{R^2}, \qquad
\Omega = r^2,
\eeq
which correspond to an AdS (dS) spacetime with $\Lambda=\mp 3/R^2$ for
the upper (lower) sign. 
The implicit solution is thus
\beq
\frac{1}{4}\left(2f - \sin 2f\right) = \frac{\pi}{2}
+ \mathcal{F}_\pm(\rho,\tilde{\kappa_1}), \label{eq:pureAdSdSsol}
\eeq
and
\beq
\mathcal{F}_\pm(\rho,\tilde{\kappa_1}) \equiv
\frac{c_s R^6}{6c_6}\bigg[
\pm\frac{\rho^4}{4} - \frac{\rho^2}{2} 
\mp 3\tilde{\kappa}_1 \rho
\pm \frac{1}{2}\log\left(1 \pm \rho^2\right)
\pm 3\tilde{\kappa}_1 A_\pm(\rho)
\bigg], \label{eq:F}
\eeq
where we have defined the dimensionless coordinate $\rho\equiv r/R$,
$\tilde{\kappa}_1\equiv\kappa_1/R^3$, as well as the function 
\beq
A_\pm(\rho) \equiv \left\{
\begin{array}{ll}
\arctan\rho,\qquad & +\\
\arctanh\rho, & -
\end{array}\right.. \label{eq:Apmdef}
\eeq
$R$ is often called the AdS (or dS) radius and as mentioned above, it
is related to the intrinsic curvature of the space time. For
convenience, we will use only the dimensionless coordinate from now
on. 

The baryon-charge density for this solution is
\beq
\mathcal{B}^r = -\frac{c_s R^3}{3\pi c_6}
\frac{\rho^3 - 3\tilde{\kappa}_1}{1 \pm \rho^2},
\eeq
and the energy- and pressure densities are given by
Eq.~\eqref{eq:specialpot_energy_pressure} with
\beq
T = \frac{c_s^2 R^6}{36c_6}
  \frac{(\rho^3 - 3\tilde{\kappa}_1)^2}{1 \pm \rho^2}, \qquad
V = \frac{c_s\pi}{2} + c_s\mathcal{F}_\pm(\rho,\tilde{\kappa}_1). 
\eeq

In order to understand the solution better, let us expand the
right-hand side of Eq.~\eqref{eq:pureAdSdSsol}, 
\beq
\frac{1}{4}\left(2f - \sin 2f\right) = \frac{\pi}{2}
-\frac{c_s\tilde{\kappa}_1}{6c_6} \rho^3 + \mathcal{O}(\rho^5).
\label{eq:AdSdSexpansion}
\eeq
This means that for both AdS and dS, $\tilde{\kappa}_1>0$ is
necessary.
Physically, this means that Skyrmion solutions must have a
nonvanishing positive baryon-charge density at the origin, which is 
exactly what one would expect.

\begin{figure}[!tpb]
\begin{center}
\mbox{
\subfloat[AdS]{\includegraphics[width=0.49\linewidth]{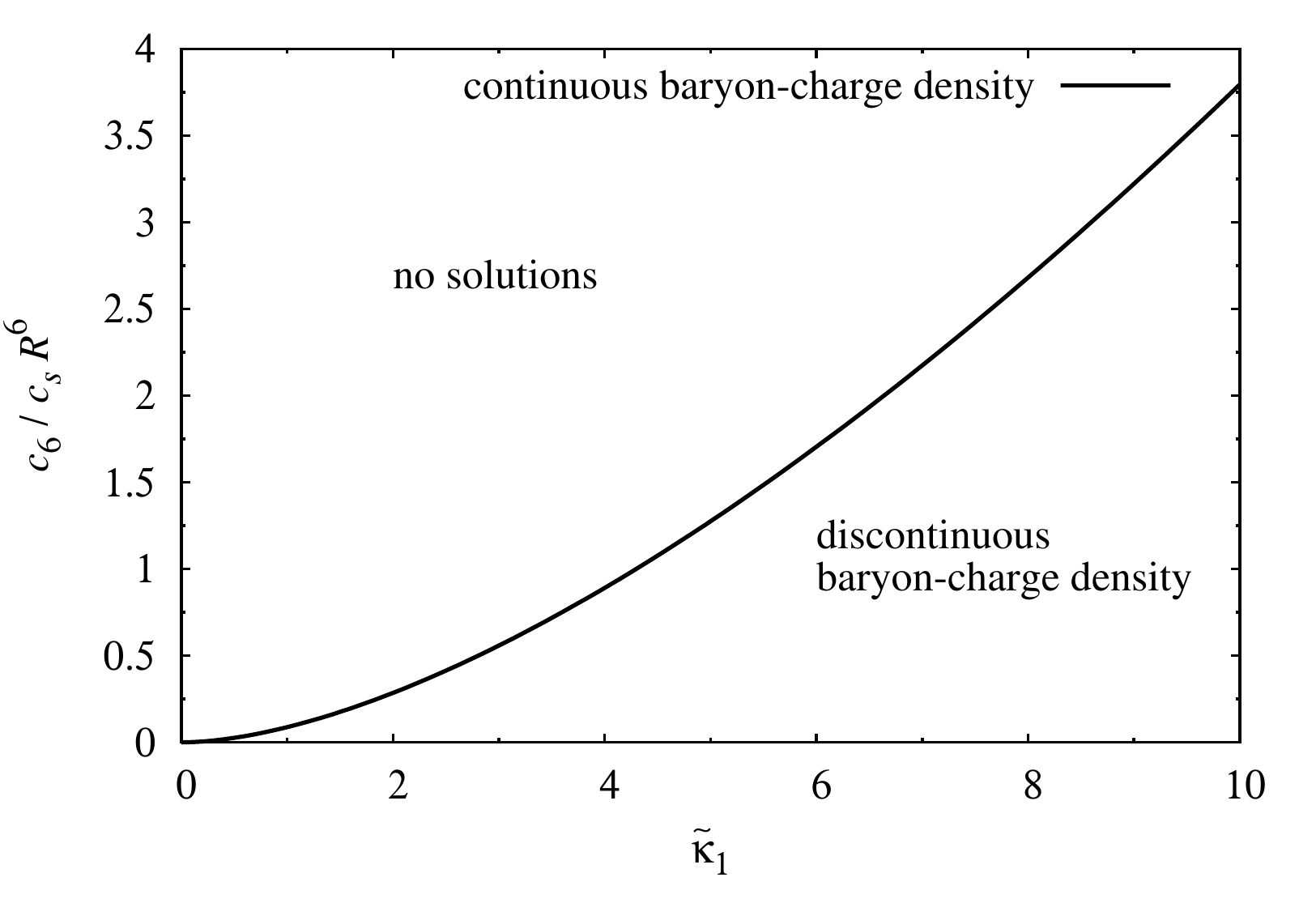}}
\subfloat[dS]{\includegraphics[width=0.49\linewidth]{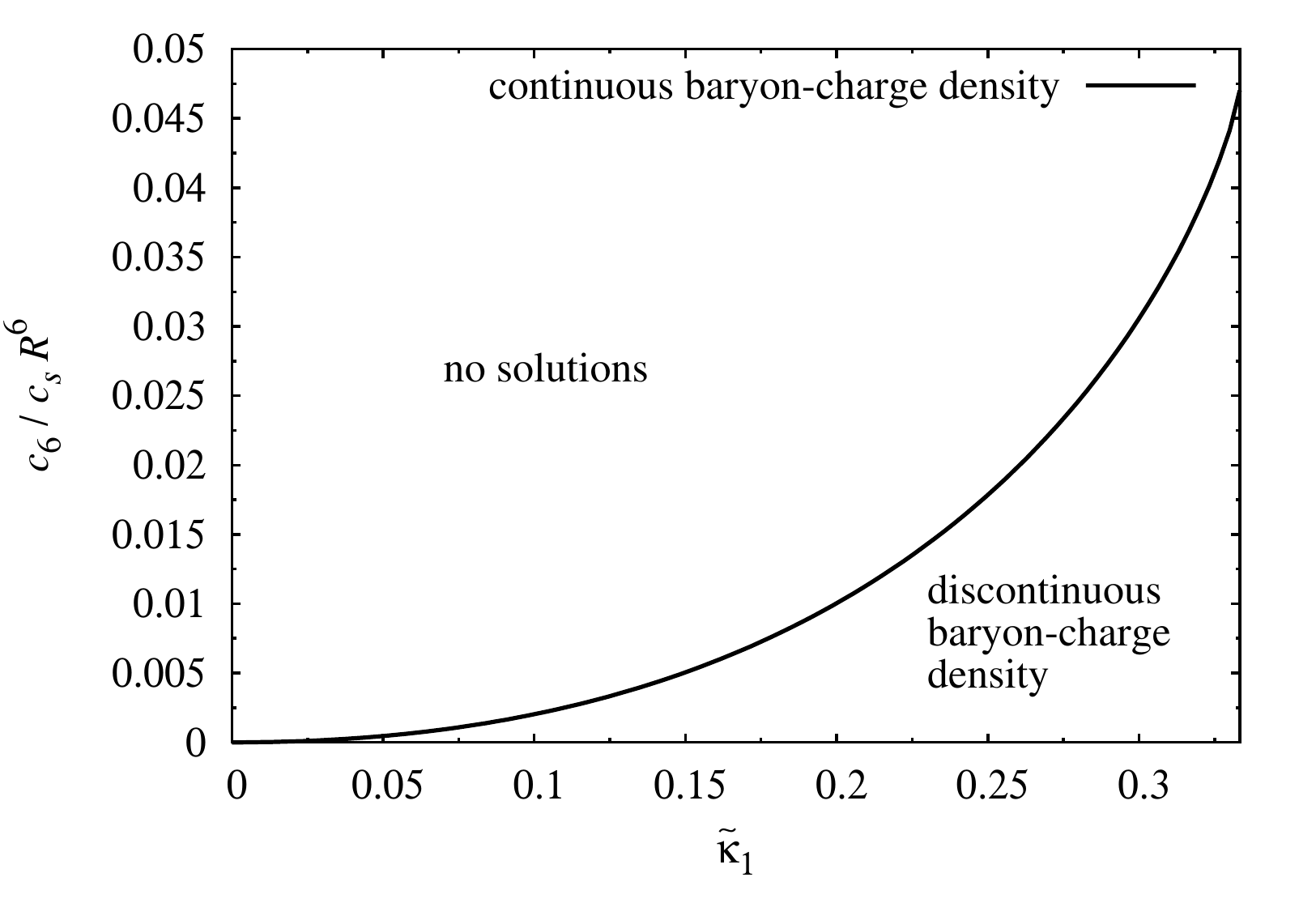}}}
\caption{Constraint Eqs.~\eqref{eq:AdScondition}
and \eqref{eq:dScondition} for having solutions on (a) AdS and (b)
dS. The line represents solutions which have continuous baryon-charge
and energy densities. } 
\label{fig:kappa1tilde}
\end{center}
\end{figure}

Let us first consider the case of AdS and hence the upper signs in the
above equations. The function $\mathcal{F}_+$ of Eq.~\eqref{eq:F},
goes to infinity for $\rho\to\infty$. Therefore, there is a lower
bound on $\tilde{\kappa}_1$ in order for the Skyrmion to satisfy the
boundary condition at the compacton radius $f(L)=0$ and thus contain
one unit of baryon charge. 
It is easy to show that there is a local minimum of $\mathcal{F}_+$
for $\tilde{\kappa}_1>0$ at $\rho=\sqrt[3]{3\tilde{\kappa}_1}$ and so
by plugging this value of $\rho$ into the
solution \eqref{eq:pureAdSdSsol}, we obtain the following implicit
lower bound for $\tilde{\kappa}_1$ 
\beq
-\frac{\pi c_6}{2c_s R^6}
+\frac{(3\tilde{\kappa}_1)^{\frac{2}{3}}}{12}
+\frac{(3\tilde{\kappa}_1)^{\frac{4}{3}}}{8}
-\frac{1}{2}\tilde{\kappa}_1 A_+[(3\tilde{\kappa}_1)^{\frac{1}{3}}]
-\frac{1}{12}\log\left[1+(3\tilde{\kappa}_1)^{\frac{2}{3}}\right]
\geq 0. \label{eq:AdScondition}
\eeq
When the above inequality is satisfied, then $\tilde{\kappa}_1$ is big
enough so that the Skyrmion profile function can flow to zero.
If, however, the inequality is saturated, i.e.~the expression is zero,
then the point where $f=0$ is coincident with where $\mathcal{B}^r=0$,
which is $f(L)=0$ with $L=\sqrt[3]{3\tilde{\kappa}_1}$ and
$\tilde{\kappa}_1$ is the solution to Eq.~\eqref{eq:AdScondition}
equaling zero.
The saturated inequality is thus the condition for the baryon-charge
density and hence the energy density to be continuous at the compacton
radius. 
Fig.~\ref{fig:kappa1tilde}a shows the constraint given in
Eq.~\eqref{eq:AdScondition} and the line represents solutions which
have continuous (but not differentiable) baryon-charge and energy
densities.

Let us now consider dS, for which the radial coordinate is only
defined in the range $\rho\in[0,1)$; i.e.~there is a cosmological
horizon at $\rho=1$ (and in turn an upper limit on the size of the
black hole \cite{Nariai:1950}). 
Proceeding as in the case of AdS, we find that there is again a local
minimum of $\mathcal{F}_-$ at $\rho=\sqrt[3]{3\tilde{\kappa}_1}$.
But since there is now a cosmological horizon at $\rho=1$, we need to
require that $\tilde{\kappa}_1<1/3$ in the following expression
\beq
-\frac{\pi c_6}{2c_6 R^3}
+\frac{(3\tilde{\kappa}_1)^{\frac{2}{3}}}{12}
-\frac{(3\tilde{\kappa}_1)^{\frac{4}{3}}}{8}
+\frac{1}{2}\tilde{\kappa}_1 A_-[(3\tilde{\kappa}_1)^{\frac{1}{3}}]
+\frac{1}{12}\log\left[1-(3\tilde{\kappa}_1)^{\frac{2}{3}}\right]
\geq 0, \label{eq:dScondition}
\eeq
where we have plugged $\rho=\sqrt[3]{3\tilde{\kappa}_1}$ into the
right-hand side of Eq.~\eqref{eq:pureAdSdSsol} and picked the lower
signs. 

If $\tilde{\kappa}_1>1/3$, $\mathcal{F}_-$ has no local minimum in the
range $\rho\in[0,1)$. Whereas $\mathcal{F}_-\to +\infty$ for
$\rho\to 1$ when $\tilde{\kappa}_1<1/3$, the divergence changes sign
for $\tilde{\kappa}_1>1/3$; i.e.~$\mathcal{F}_-\to -\infty$ for
$\rho\to 1$.
We will now consider the case of the compacton size being close to
unity: $f(L)=0$ with $L=1-\epsilon$. Expanding the right-hand side of
Eq.~\eqref{eq:pureAdSdSsol} to first order in $\epsilon$, we get
\begin{align}
\frac{\pi c_6}{2c_s R^6}
-\frac{1}{8}
-\frac{1}{12}\log 2
-\frac{1}{4}(\log 2 - 2)\tilde{\kappa}_1
+\frac{1}{12} (3\tilde{\kappa}_1 - 1) \log\epsilon
- \frac{3}{8} \epsilon (\tilde{\kappa}_1 - 1)
+ \mathcal{O}(\epsilon^2) = 0. 
\end{align}
Solving this equation -- which is the expansion in $\epsilon$ to
linear order -- gives us
\begin{align}
\epsilon = -\frac{2(1-3\tilde{\kappa}_1)}{9(1-\tilde{\kappa}_1)}
W\left[-\frac{9(1-\tilde{\kappa}_1)}{1-3\tilde{\kappa}_1} 2^{-1-\frac{1}{1-3\tilde{\kappa}_1}}
  \left(2^{-6\tilde{\kappa}_1}
  \exp\left(-3+\frac{12\pi c_6}{c_s R^6}+12\tilde{\kappa}_1\right)
  \right)^{\frac{1}{2(1-3\tilde{\kappa}_1)}}\right],
\end{align}
where $W$ is the Lambert $W$-function, which is the inverse function
of
\beq
F(W) = W e^W.
\eeq
Although the profile function $f$ goes to zero at $\rho=1-\epsilon$
for $\tilde{\kappa}_1>1/3$, the baryon-charge density will be finite
at that radius and thus be discontinuous. In turn the energy density
will be discontinuous.

\begin{figure}[!thp]
\begin{center}
\mbox{
\subfloat[]{\includegraphics[width=0.49\linewidth]{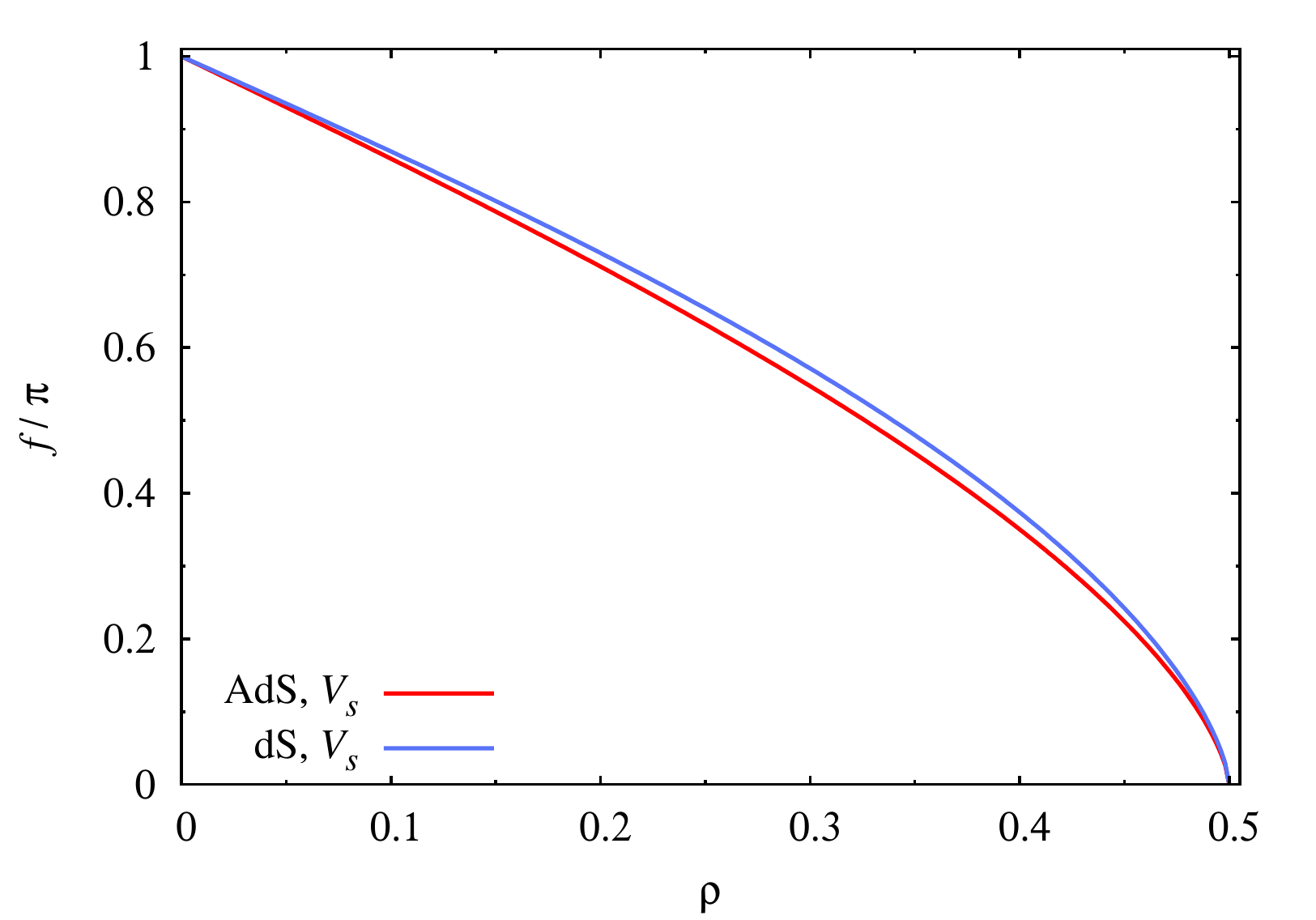}}\
\subfloat[]{\includegraphics[width=0.49\linewidth]{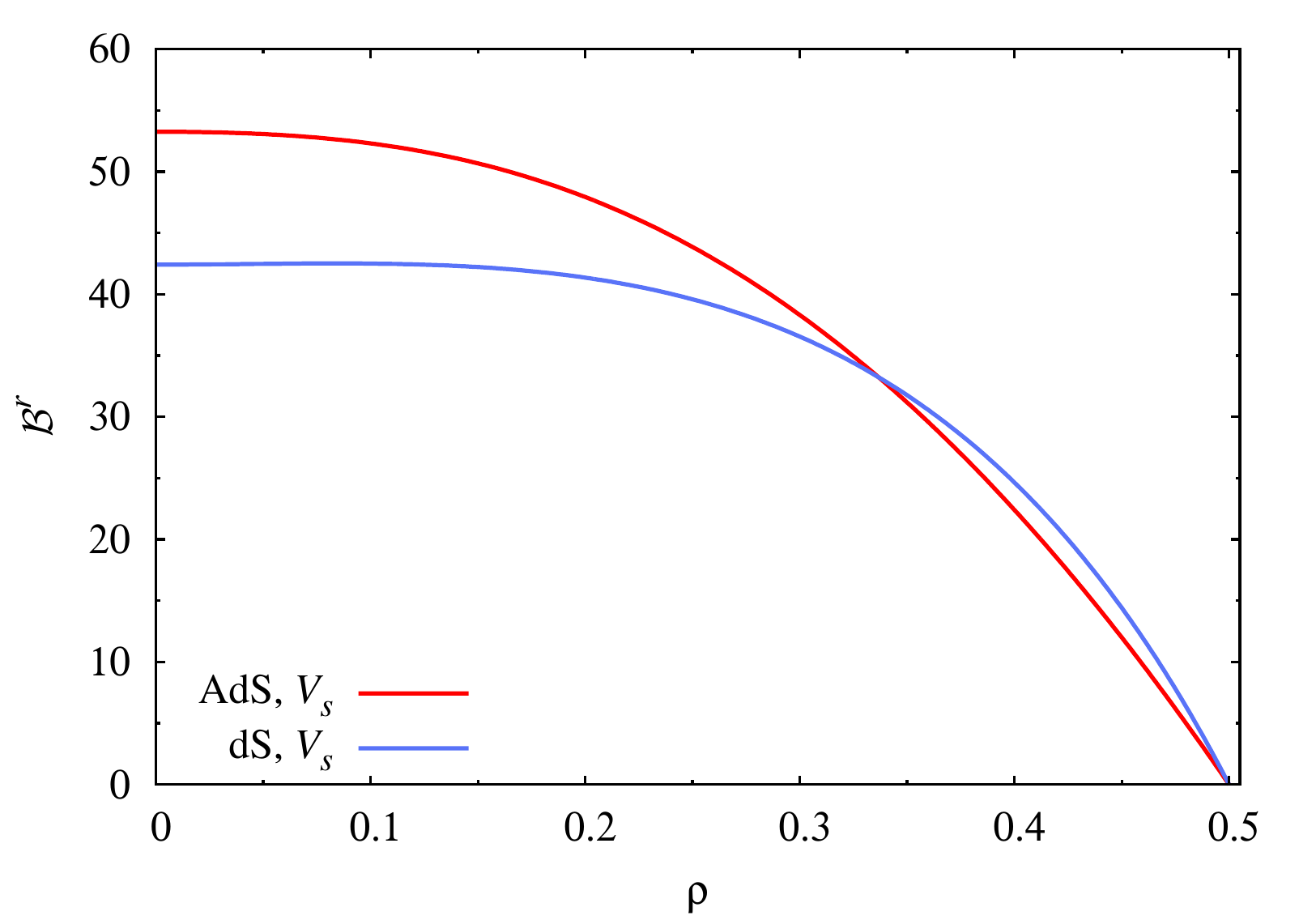}}}
\mbox{
\subfloat[]{\includegraphics[width=0.49\linewidth]{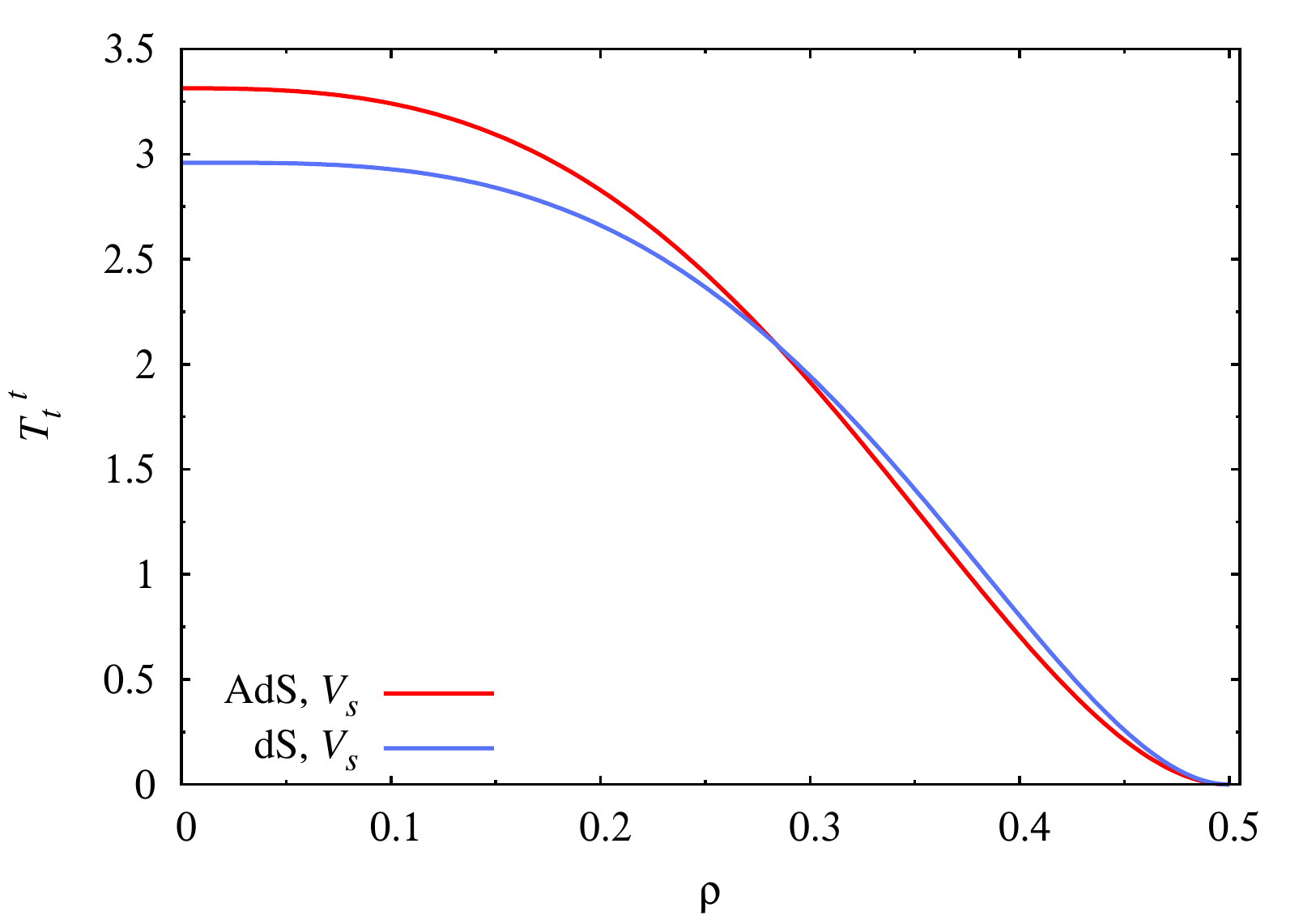}}\
\subfloat[]{\includegraphics[width=0.49\linewidth]{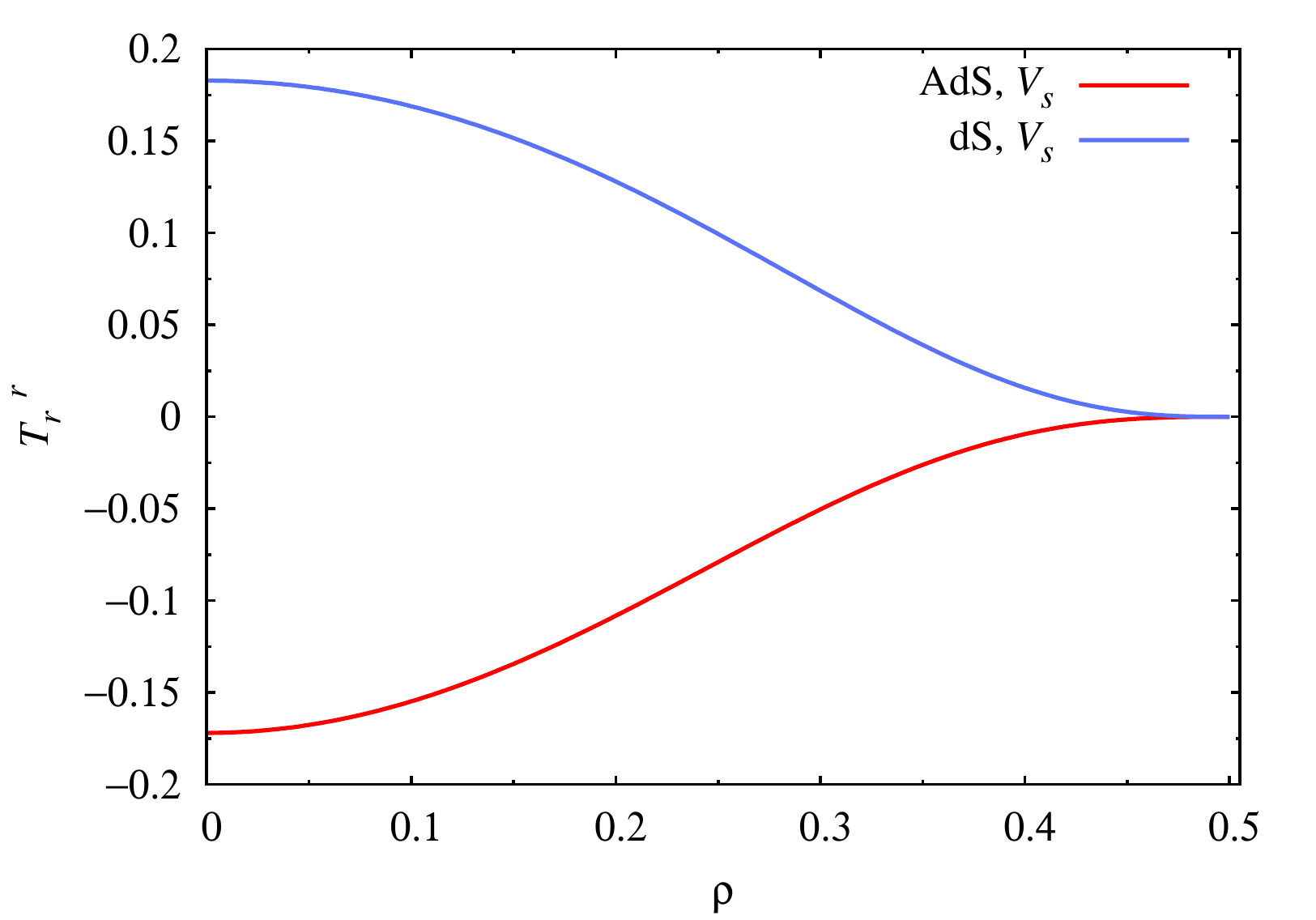}}}
\caption{(a): Profile functions, $f$, (b) baryon-charge densities, (c)
  energy densities and (d) pressure densities, for the analytic
  (non-BPS) Skyrmions, with the special potential in the pure AdS and
  dS spacetime backgrounds without backreaction ($\alpha=0$). 
  We have chosen the value of
  $\tilde{\kappa}_1=\tfrac{1}{3\times 2^3}=1/24$ and in turn set
  $c_6=[11-8\arccot 2 - 32\log(5/4)]/(192\pi)\simeq 2.49053\times 10^{-4}$
  for the AdS case and 
  $c_6=[5-64\log 2 + 36\log 3]/(192\pi)\simeq 3.12711\times 10^{-4}$ for
  the dS case, so that the baryon-charge and energy densities are
  continuous functions at the compacton radius. 
  The constants are set as $c_{s}=R=1$.
}
\label{fig:cbgadsds}
\end{center}
\end{figure}

Solutions to Eq.~\eqref{eq:pureAdSdSsol}, saturating the
condition \eqref{eq:AdScondition} [\eqref{eq:dScondition}], which are
(non-BPS) Skyrmions with the special potential \eqref{eq:specialpot}
in the pure AdS [dS] background without taking into account the
backreaction onto gravity; i.e.~in the limit of $\alpha=0$, are shown
in Fig.~\ref{fig:cbgadsds}.

It is interesting to note that the baryon-charge density has a peak in
the case of dS, while it does not in the case of AdS. More concretely,
a peak in the baryon-charge density occurs if a real zero of the
second derivative of the baryon-charge density is smaller than the
compacton size, i.e.~if the following condition is satisfied
\begin{align}
\pm\Xi
\pm\frac{9\tilde{\kappa}_1^2\pm 1}{\Xi}
\mp3\tilde{\kappa}_1
< L = (3\tilde{\kappa}_1)^{\frac{1}{3}}, \qquad
\Xi \equiv \sqrt[3]{-27\tilde{\kappa}_1^3 \mp 3\tilde{\kappa}_1
  +\sqrt{-18\tilde{\kappa}_1^2\mp 1\mp 81\tilde{\kappa}_1^4}}.
\end{align}

Notice also that the pressure density of the solution in AdS [dS] is 
negative [positive] up to the compacton radius where it vanishes; this
compensates the intrinsic curvature of the spacetime.

\subsection{The gravitating Skyrmion with special potential}
\label{sec:gravsk}

In this section we turn on the gravitational coupling $\alpha>0$ and
consider the system fully coupled to the gravitational background.
In order to simplify the problem, we will take the metric to be in the
form
\beq
ds^2 = e^{-2\lambda} C dt^2 - C^{-1} dr^2
- r^2\left(d\theta^2 + \sin^2\theta d\phi^2\right),
\eeq
which means that $N=e^{-\lambda}$, and $\Omega = r^2$, where
$\lambda=\lambda(r)$ and $C=C(r)$ are radial functions. 
Since we are still considering the special
potential \eqref{eq:specialpot}, Eq.~\eqref{eq:specialsolf} is still a
solution to the matter equation of motion. However, now we also have 
to solve the inhomogeneous Einstein equations, which boil down to
\begin{align}
&r\lambda' + \frac{2\alpha c_6\sin^4(f) f_r^2}{r^2} = 0, \\
&-1 + C + r C' + r^2\Lambda
+C\frac{2\alpha c_6\sin^4(f) f_r^2}{r^2} 
+\frac{1}{2}\alpha c_s r^2\left(2f-\sin 2f\right) = 0.
\end{align}
By inserting Eqs.~\eqref{eq:specialsolf} and \eqref{eq:B0specialsol}
into the above system, we can eliminate the Skyrmion profile function
and we have the following integro-differential equations
\beq
&&\lambda'
+\frac{\alpha c_s^2r}{2c_6 e^{-2\lambda}C^2}\left(
\int_{r_0}^r dr'\; {r'}^2 e^{-\lambda} - \kappa_1\right)^2 = 0,
\label{eq:integrodiff1}\\
&&-1 + C + r C' + r^2\Lambda 
+\frac{\alpha c_s^2 r}{2c_6 e^{-2\lambda}C}\left(
\int_{r_0}^r dr'\; {r'}^2 e^{-\lambda} - \kappa_1\right)^2 \non
&&\quad
+\frac{\alpha r^2 c_s^2}{c_6}\int_{r_0}^r
 dr' \; \frac{{r'}^2}{e^{-\lambda}C}
 \left(\int_{r_0}^{r'} dr''\; {r''}^2 e^{-\lambda} - \kappa_1\right) 
+\frac{1}{2}\alpha c_s r^2\left(2f_0 - \sin 2f_0\right) = 0.
\label{eq:integrodiff2}
\eeq
The first observation, which fits nicely with the results of
section \ref{sec:skcurvbg}, is that when $\alpha>0$ is turned on, then
$N=e^{-\lambda}$ becomes a nontrivial function (as opposed to the case
of $\alpha=0$ where $N$ is simply unity).

In principle, one can pick a trial function for $\lambda$ and
determine $C$ from Eq.~\eqref{eq:integrodiff1}, which then has to
satisfy the remaining Einstein equation, \eqref{eq:integrodiff2}.
Unfortunately, we have not been able to find an analytic solution to
this coupled system of integro-differential equations and we therefore
turn to numerical methods for finding gravitating Skyrmion solutions
(with $\alpha>0$).

We use a fourth-order Runge-Kutta method and implement a simple
trapezoidal integration method in the loop that integrates the two
integrals needed for Eqs.~\eqref{eq:integrodiff1}
and \eqref{eq:integrodiff2}. We choose the step size of the numerical
integration to be $\Delta\rho=5\times 10^{-4}$. 
Since the gravitational backreaction induces a non-constant $\lambda$, 
we need to adjust $\tilde{\kappa}_1$ in order to make the
baryon-charge density vanish at the radius of the
compacton. Similarly, we adjust the BPS Skyrme term coefficient,
$c_6$, such that the Skyrmion profile function, $f$, vanishes at the
same radius as that without gravitational backreaction (the
corresponding $\alpha=0$ solution).
The method we use to adjust the coefficients $\tilde{\kappa}_1$ and
$c_6$ is to begin the above-described Runge-Kutta integration with the
initial guess of the solution without gravitational backreaction and
then calculate the following error
\beq
e = |f(\ell)| + |\mathcal{B}^r(\ell)|, \label{eq:e}
\eeq
where $\ell$ is either the zero of $f$ or the minimum value (the
turning point). If $\tilde{\kappa}_1$ is too big, $f$ will have a
zero, but if it is too small, it will simple turn around with a cusp
and increase monotonically from that radius on.
Finally, we employ a steepest descent algorithm on the $e$
function \eqref{eq:e} with the variables $(\tilde{\kappa}_1,c_6)$.

\begin{figure}[!p]
\begin{center}
\mbox{
\subfloat[]{\includegraphics[width=0.49\linewidth]{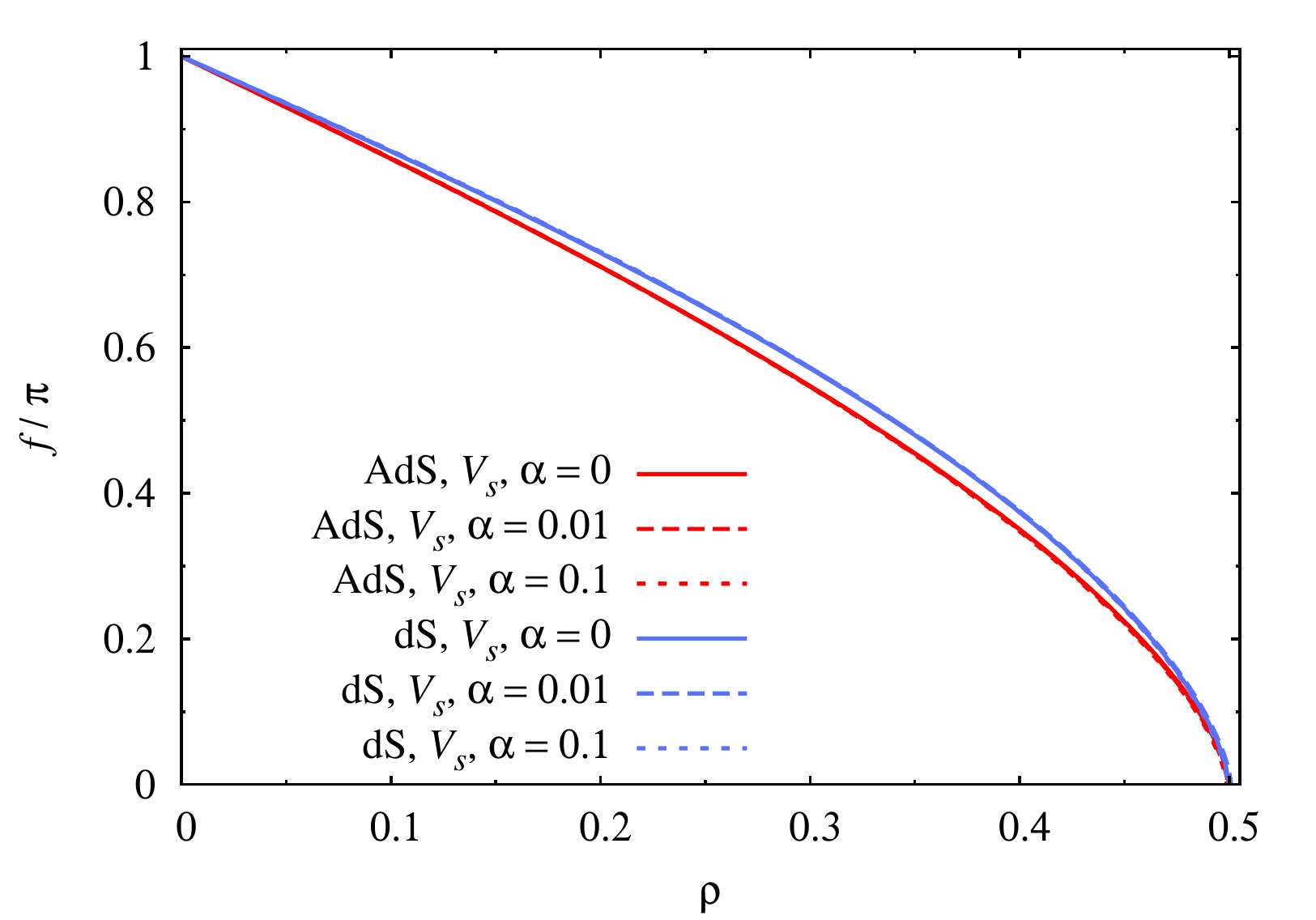}}\
\subfloat[]{\includegraphics[width=0.49\linewidth]{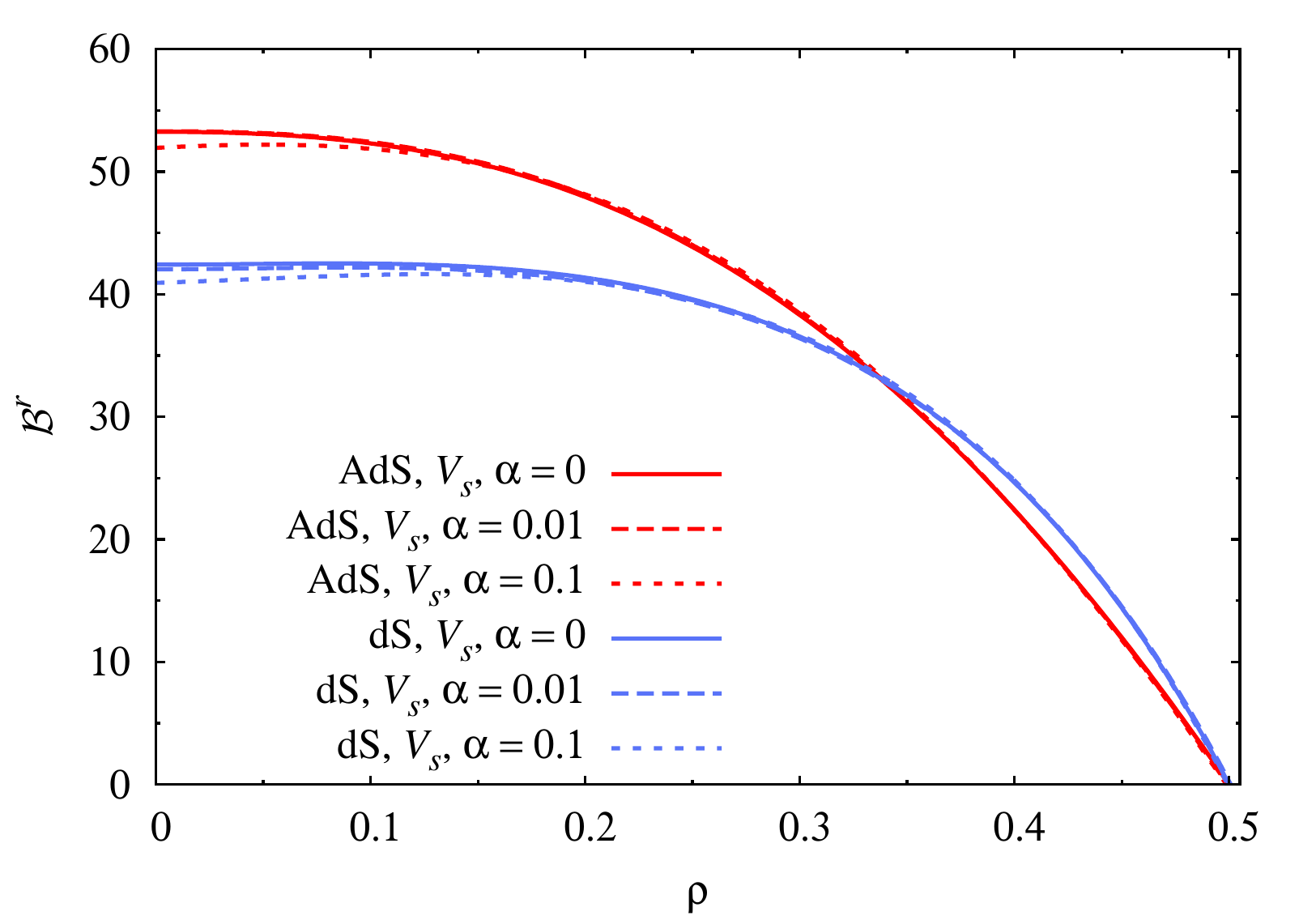}}}
\mbox{
\subfloat[]{\includegraphics[width=0.49\linewidth]{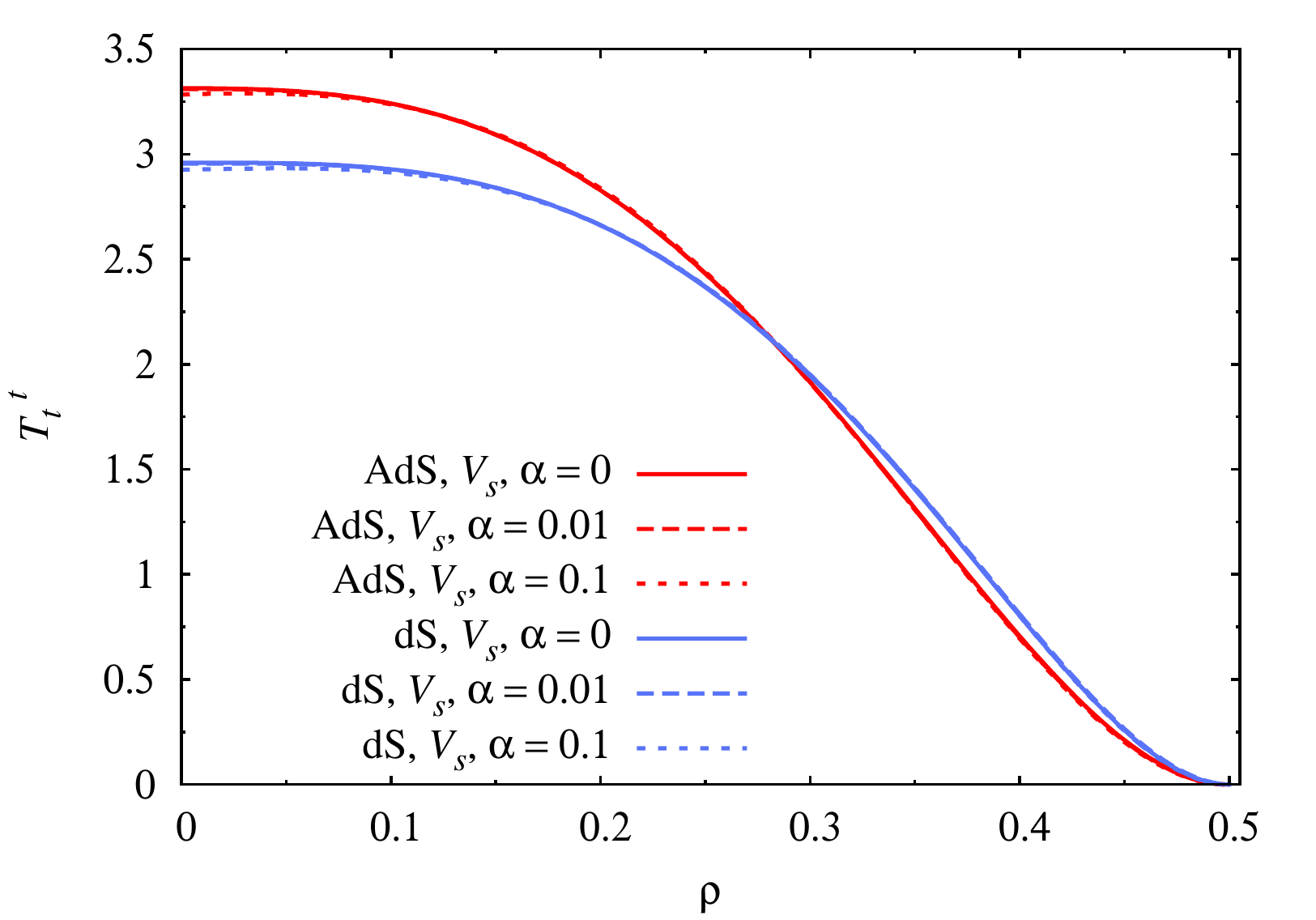}}\
\subfloat[]{\includegraphics[width=0.49\linewidth]{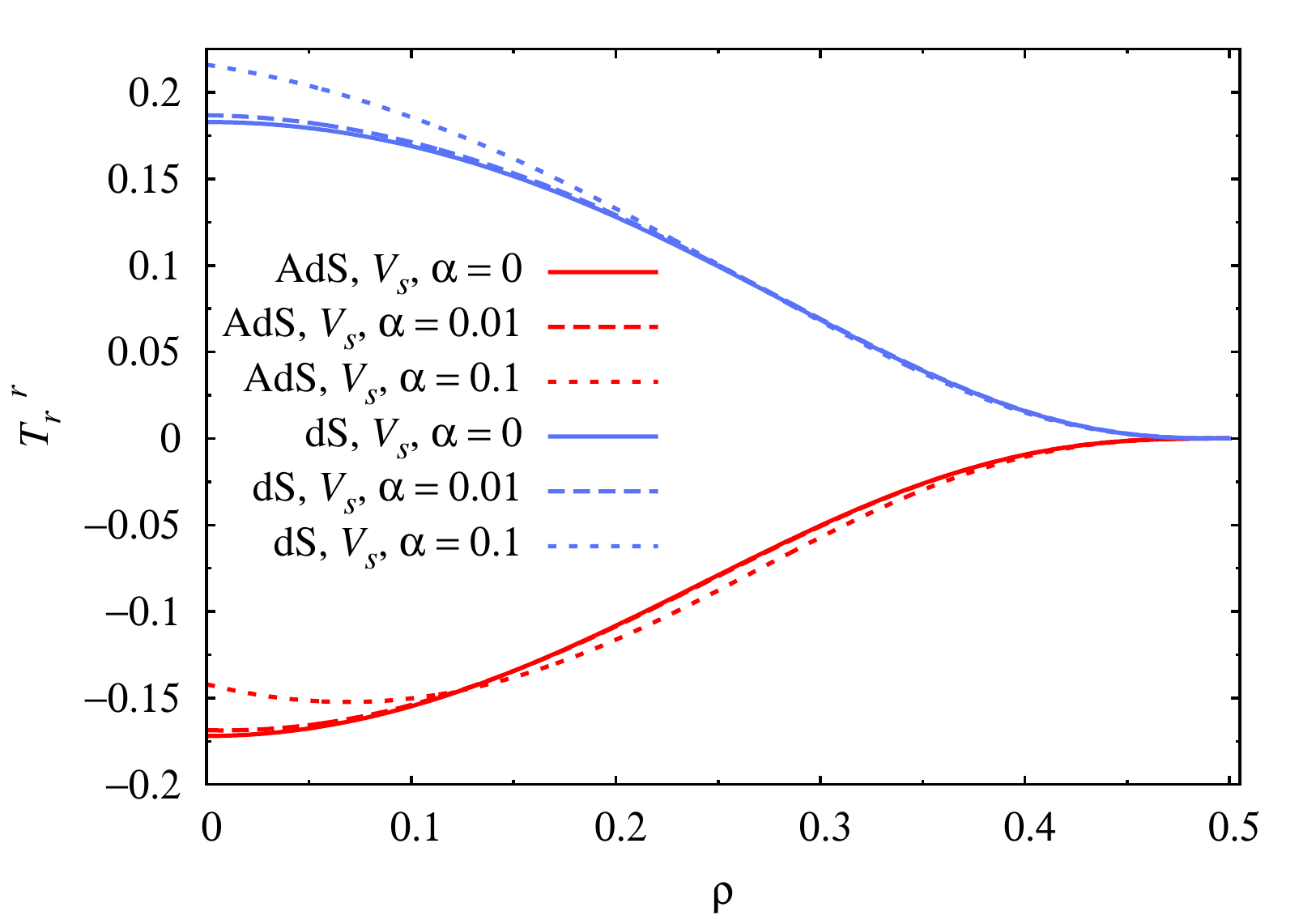}}}
\mbox{
\subfloat[]{\includegraphics[width=0.49\linewidth]{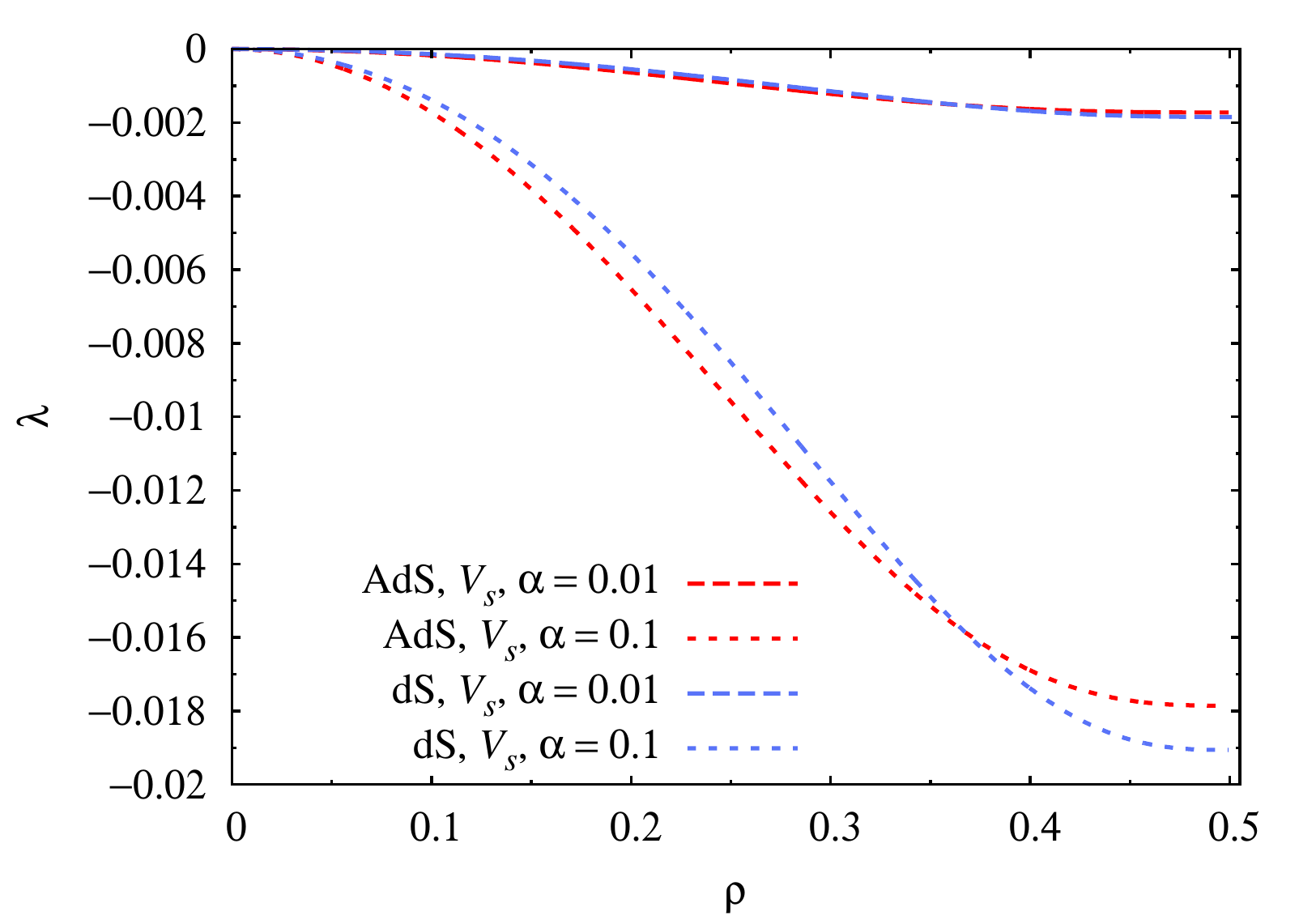}}
\subfloat[]{\includegraphics[width=0.49\linewidth]{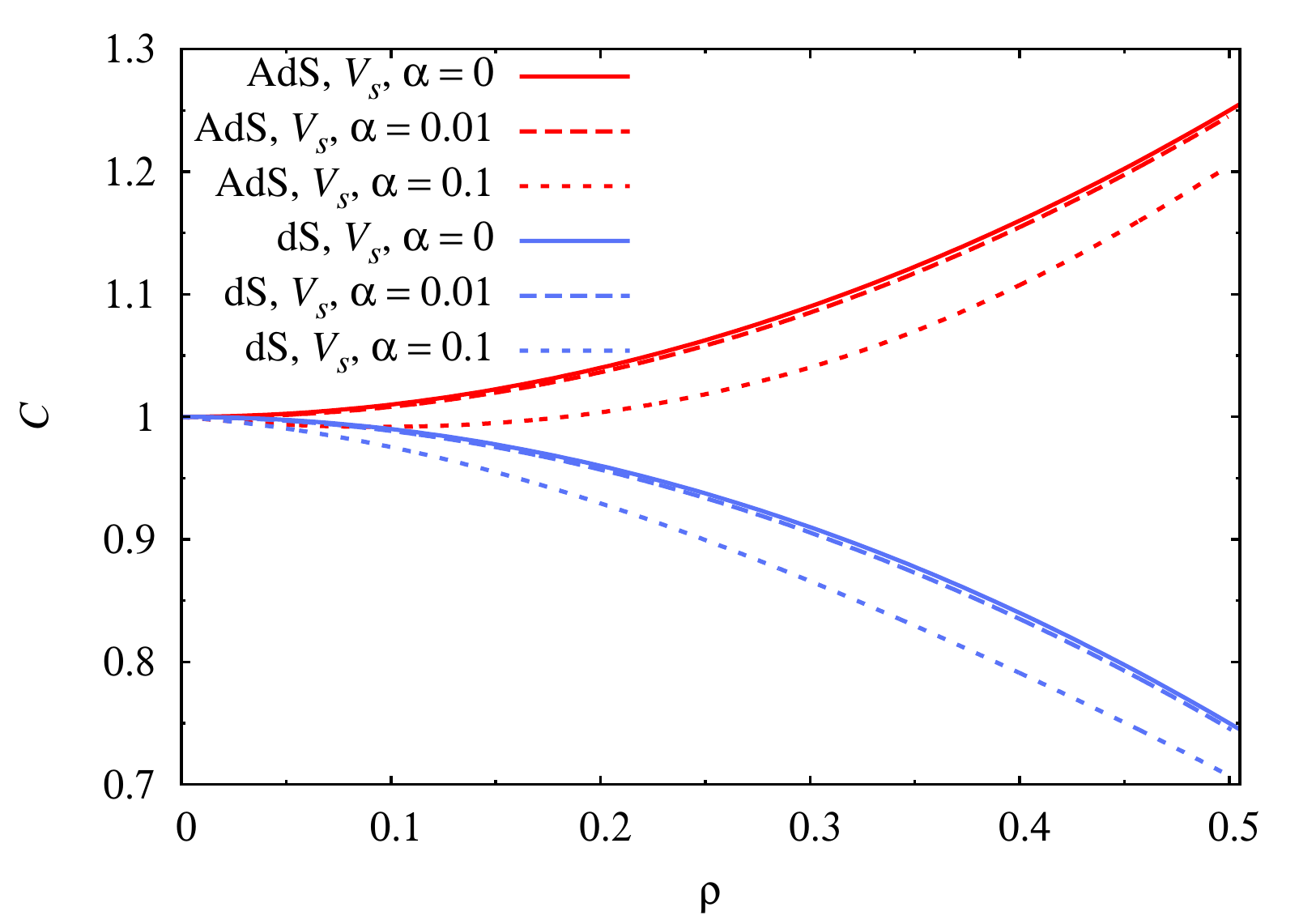}}}
\caption{(a): Profile functions, $f$, (b) baryon-charge densities, (c)
  energy densities, (d) pressure densities and metric functions (e)
  $\lambda$ and (f) $C$, for numerical (non-BPS) Skyrmions, with the
  special potential in AdS and dS backgrounds with backreaction
  $\alpha=0.01,0.1$ and without $\alpha=0$. 
  The constants are set as $c_{s}=R=1$. 
  The numerically integrated baryon charges are 
  $B_{{\rm AdS},V_s,\alpha=0.01}=0.999999990$,
  $B_{{\rm AdS},V_s,\alpha=0.1}=0.999999981$,
  $B_{{\rm dS},V_s,\alpha=0.01}=0.999995222$ and
  $B_{{\rm dS},V_s,\alpha=0.1}=1.00000109$.
}
\label{fig:ide}
\end{center}
\end{figure}

Fig.~\ref{fig:ide} shows numerical solutions for the Skyrmions in
anti-de Sitter and de Sitter backgrounds.
For comparison, we show the analytic solutions alongside the numerical
ones (the $\alpha=0$ ones), except for $\lambda$ which vanishes for
$\alpha=0$. 
As one would expect, the numerical solutions with small $\alpha$ are
very well approximated by the analytic solutions; in particular the 
$\alpha=0.01$ ones.
The values of the parameters, $\tilde{\kappa}_1$ and $c_6$ for the
solutions of Fig.~\ref{fig:ide} are shown in Tab.~\ref{tab:ide}.

\begin{table}[!th]
\begin{center}
\caption{Parameters $\tilde{\kappa}_1$ and $c_6$ for the numerical
Skyrmion solutions in AdS and dS with gravitational couplings
$\alpha=0,0.01,0.1$. }
\label{tab:ide}
\begin{tabular}{l|cr@{\,}cr@{\,}c}
\multirow{4}{*}{AdS} & 
$\alpha$ & & $\tilde{\kappa}_1$ & & $c_6$\\
\cline{2-6}
& 0.00 & $1/24\simeq$ & 0.0416667 &
  $[11-8\arccot 2 - 32\log\frac{5}{4}]/(192\pi)\simeq$ & 0.000249053\\
& 0.01 & & 0.0415637 & & 0.000248301\\
& 0.10 & & 0.0419864 & & 0.000257302\\
\hline\hline
\multirow{4}{*}{dS} & 
$\alpha$ & & $\tilde{\kappa}_1$ & & $c_6$\\
\cline{2-6}
& 0.00 & $1/24\simeq$ & 0.0416667 &
  $[5-64\log 2 + 36\log 3]/(192\pi)\simeq$ & 0.000312711\\
& 0.01 & & 0.0419284 & & 0.000317556\\
& 0.10 & & 0.0421486 & & 0.000327807
\end{tabular}
\end{center}
\end{table}

\section{Discussion and conclusion}\label{sec:discussion}

The BPS-ness of Skyrmions in the BPS Skyrme model requires a constant 
time-time component of the metric or trivial warp factor. It does
however allow for a curved 3-space and we have thus found analytic BPS 
Skyrme solutions on the 3-sphere and on the 3-hyperboloid. The BPS
solutions can be found for a large class of potentials.
These solutions are pressure-less and hence they solve the spatial
part of the Einstein equations; they are solutions on a curved space
(but not curved spacetime).
For the 3-sphere, however, we found a solution to all the Einstein
equations, for a constant potential, by fixing the curvature of the
3-sphere (the cosmological constant) in terms of the gravitational
coupling multiplied by the potential parameter.
In order for the Skyrmion to be BPS and solve the time-time component
of the Einstein equations, only a constant potential is allowed.
This solution is thus unique. 

The pursuit of gravitating Skyrmions on a curved spacetime
background (as opposed to a curved spatial background) in the BPS
Skyrme model, however, turned out to be easier if a special potential
is chosen.
For this particular potential, we have been able to find analytic
solutions -- which however are non-BPS -- on both the anti-de Sitter
and the de Sitter spacetime backgrounds by neglecting backreaction
to gravity.
These solutions are analytic solutions in the limit of vanishing
gravitational coupling. Once we turn back on the gravitational
coupling, $\alpha>0$, the governing system of equations becomes a
coupled set of two integro-differential equations for which we have
not been able to find analytic solutions. 
We have, however, found numerical solutions to this system and shown
that for weak gravitational coupling, the analytic solution is a good
approximation.

One curiosity has arisen from this study, namely that with the chosen
special potential \eqref{eq:specialpot}, we have not been able to find
a regular Skyrmion solution with a black hole horizon.
This is in contrast to the black holes with Skyrme
hair \cite{Luckock:1986tr,Droz:1991cx,Bizon:1992gb,Shiiki:2005pb} that
were the inspiration for this work. In the Einstein-Skyrme theory, 
regularity of the Skyrmion solution at the horizon dictates the first
derivative of the Skyrmion profile function. Logically, there are two 
possibilities; when a horizon is present, either no regular solution
exists for the special potential or no regular solution exists at all
in the BPS Skyrme model. One way to approach this question is to turn
on a kinetic term (and possibly the Skyrme term) and take the limit
where their coefficients vanish. That is beyond the scope of this
paper though.

As we mentioned in the end of Sec.~\ref{sec:model}, the modification
of the BPS equation to include a constant pressure, as for instance
considered in Refs.~\cite{Adam:2014nba,Adam:2015lpa}, is not enough
for obtaining a first-order equation (BPS-like equation) that solves
the full second-order equation of motion on a curved spacetime
background. It would be very interesting to find some way or limit in
which one may obtain a first-order equation for Skyrmions on a curved
spacetime background (other than the special potential that we have
considered in this paper). That is, however, beyond the scope of this
paper.

In this paper, we have considered numerical solutions only for small
values of $\alpha$. It is expected that there is critical value,
$\alpha_{\rm cr}$, beyond which no Skyrmion
exists \cite{Bizon:1992gb,Bizon:1998kq}. 
We leave this study for the future. 

The quadratic potential in Eq.~\eqref{eq:pot} possesses two discrete 
vacua and hence admits a domain wall interpolating between
them \cite{Nitta:2012wi,Nitta:2012rq,Gudnason:2013qba,Gudnason:2014nba,Gudnason:2014gla,Gudnason:2014hsa}. 
Lumps inside the domain wall are Skyrmions from the bulk 
point of view that we called domain wall
Skyrmions \cite{Nitta:2012wi,Nitta:2012rq,Gudnason:2014nba,Gudnason:2014gla,Gudnason:2014hsa}. 
Gravitational domain-wall Skyrmions in the BPS Skyrme model 
could also be an interesting future problem.

In this paper we have considered only Skyrmions of charge one. In the
literature, axially symmetric Skyrmions are well
known \cite{Kopeliovich:1987bt,Manton:1987xf,Verbaarschot:1987au} and 
their gravitational counterparts, i.e.~the axially symmetric
Skyrmion-black hole systems were considered for the normal Skyrme
model coupled to gravity in Refs.~\cite{Sawado:2004yq,Sato:2006xy}
(see also Ref.~\cite{Shnir:2015aba}). 
This can also be considered in the model considered in this paper.
In the case of the BPS solutions with neglected backreaction to
gravity, a very simple extension to higher $B$ is possible by
increasing the axial winding
$\pi^1+i\pi^2=e^{i\phi}\to\pi^1+i\pi^2=e^{i B\phi}$, but keeping
spherical symmetry for the Skyrmion profile function as in
Ref.~\cite{Adam:2010fg}. This is possible because of the
volume-preserving diffeomorphism of the BPS
system \cite{Adam:2010fg}.
When the system is not BPS or finite coupling to gravity is
considered, then the situation is more complicated. This may be
considered in future studies.

\subsection*{Acknowledgments}

S.~B.~G.~would like to thank Andrzej Wereszczynski and Christoph Adam 
for comments on the draft and Roberto Auzzi for discussions.
S.~B.~G.~also thanks the Recruitment Program of High-end Foreign
Experts for support.
The work of M.~N.~is supported in part by a Grant-in-Aid for
Scientific Research on Innovative Areas ``Topological Materials
Science'' (KAKENHI Grant No.~15H05855) and ``Nuclear Matter in Neutron
Stars Investigated by Experiments and Astronomical Observations''
(KAKENHI Grant No.~15H00841) from the the Ministry of Education,
Culture, Sports, Science (MEXT) of Japan. The work of M.~N.~is also
supported in part by the Japan Society for the Promotion of Science
(JSPS) Grant-in-Aid for Scientific Research (KAKENHI Grant
No.~25400268) and by the MEXT-Supported Program for the Strategic
Research Foundation at Private Universities ``Topological Science''
(Grant No.~S1511006).

\end{document}